\documentclass[prd,letterpaper,superscriptaddress,reprint,nofootinbib]{elsarticle} 
\usepackage{lineno}
\usepackage{url}
\usepackage{hyperref}
\usepackage{color}
\include{rlFig}
\begin{document}
\begin{frontmatter}
\title{Triboelectric Backgrounds to Radio-based Polar UHE Neutrino Experiments}


\author[ulb]{J.~A.~Aguilar}
\author[uci]{A.~Anker}
\author[osu]{P.~Allison}
\author[chiba]{S.~Archambault}
\author[ucib]{P.~Baldi}
\author[uci]{S.~W.~Barwick}
\author[osu]{J.~J.~Beatty}
\author[UppPh]{J.~Beise}
\author[ku]{D.~Besson}
\author[wipac]{A.~Bishop}
% ...

\address[ulb]{Universite Libre de Bruxelles, Science Faculty CP230, B-1050 Brussels, Belgium}
\address[uci]{Department of Physics and Astronomy, University of California, Irvine, CA 92697, USA}
\address[osu]{Dept. of Physics, Center for Cosmology and AstroParticle Physics, The Ohio State University, Columbus, OH 43210}
\address[chiba]{Dept. of Physics, Chiba University, Chiba, Japan}
\address[ucib]{Department of Information and Computer Science, University of California, Irvine, CA 92697, USA}
\address[UppPh]{Uppsala University Department of Physics and Astronomy, Uppsala SE-752 37, Sweden}
\address[ku]{University of Kansas Department of Physics and Astronomy, Lawrence KS 66045, USA}
\address[wipac]{Dept. of Physics, University of Wisconsin-Madison, Madison,  WI 53706, USA}
% ...

\date{\today}
\begin{abstract}
The proposed IceCube-Gen2 seeks to instrument approximately 500 square kilometers of Antarctic ice near the geographic South Pole with radio antennas, in the hopes of observing the highest-energy neutrinos (E$>$1 EeV) populating the Universe. For this purpose IceCube-Gen2 will exploit the impulsive radio emission produced by neutrino interactions in the polar ice cap.
In such polar-sited radio-frequency neutrino search experiments, rare single event candidates must be unambiguously identified above backgrounds. 
Background rejection strategies to date primarily target thermal noise fluctuations and also impulsive radio-frequency signals of anthropogenic origin. 
In this paper, we consider the possibility that `fake' neutrino signals may also be generated naturally via the `triboelectric effect.'
This broadly describes any process in which force applied at a boundary layer results in displacement of surface charge, leading to the production of an electrostatic potential difference $\Delta$V. 
Wind blowing over granular surfaces such as snow can induce such a potential difference, with subsequent coronal discharge. Discharges over timescales as short as nanoseconds can then lead to radio-frequency emissions at characteristic MHz -- GHz frequencies. 

Using data from various past and current neutrino experiments, we find that such backgrounds are observed by all radio neutrino experiments and are generally characterized by: 
a) a threshold wind velocity which likely depends on the experimental signal trigger threshold and layout; for the experiments considered herein, this value is typically $\cal{O}$(10 m/s),
b) frequency spectra generally shifted to the low-end of the frequency regime to which current radio experiments are typically sensitive (100-200 MHz), 
c) for the strongest background signals, an apparent preference for discharges from above-surface structures, although the presence of more isotropic, lower amplitude triboelectric discharges cannot be excluded. 
 
\end{abstract}
\end{frontmatter}

\flushbottom

\section{Introduction}
The last few years have witnessed the emergence of so-called Multi-Messenger Astronomy (MMA)~\cite{bartos2017multimessenger}, wherein an individual source may be observed via its charged cosmic-ray, neutrino, electromagnetic and gravitational wave emissions. 
In the neutrino sector, the IceCube experiment~\cite{halzen2010invited} has made the first measurements of the diffuse extra-terrestrial neutrino flux above 1 TeV energies~\cite{Aartsen:2020aqd}, as well as a possible neutrino correlation with the gamma-ray active TXS 0506+056 blazar~\cite{icecube2018multimessenger}. 
At higher energies ($E>100$ PeV), detection of so-called `cosmogenic neutrinos'~\cite{kotera2010cosmogenic,ahlers2012minimal} would represent a major milestone in multi-messenger particle astrophysics.
Interactions of Ultra-High Energy Cosmic Rays (UHECR) at energies in excess of 10 EeV with the Cosmic Microwave Background (CMB) produce such neutrinos via photoproduction of the $\Delta(1232)$ resonance; 
such interactions also effectively remove UHECR from the astrophysical `beam', leading to the Greisen-Zatsepin-Kuzmin (GZK~\cite{GZK1,GZK2,GZK3}) maximum energy cutoff in charged particle cosmic ray spectra probed by the Auger~\cite{aab2020features} and Telescope Array~\cite{TelescopeArray:2015dcv} experiments.
Simple particle physics arguments predict three ultra-high energy neutrinos (UHEN) produced per UHE proton-CMB interaction, with neutrino energy spectra peaking approximately two decades in energy below the $E_{\mathrm{GZK}}\sim 10^{19.5}$ eV cut-off; even a non-observation of such neutrinos would help guide our understanding of the cosmic ray flux and physics at these energies. 

This has spurred a large number of experiments with the express goal of making the first-ever observation of the diffuse cosmogenic neutrino flux. 
Owing to the miniscule event rate at these energies (of order one per year per 10 cubic kilometers of sensitive neutrino target volume), all detection strategies require a large product of [target volume]$\times$[observation time]. 
Cold ($T<$230 K) polar ice provides a nearly ideal target medium. The extraordinary clarity at radio-frequencies, with attenuation lengths measured in kilometers~\cite{barwick2005south,Allison:2019rgg,kravchenko2004situ}, allow a single embedded detector to scan gigaton neutrino target volumes. Following the RICE~\cite{kravchenko2004situ,Kravchenko:2011im,RICE} (1995-2011) and AURA~\cite{landsman2007aura} experiments (2007--2011), several efforts, both current ARA~\cite{allison2012design}, ARIANNA~\cite{Barwick:2014pca}, ANITA~\cite{GorhamAllisonBarwick2009}, RNO-G~\cite{Aguilar:2020xnc} and planned radio array of IceCube Gen-2~\cite{IceCube-Gen2:2020qha} and PUEO~\cite{PUEO:2020bnn} exploit detection of the coherent long-wavelength (radio) Cherenkov emission produced by collisions of neutrinos, of any flavor, with cold polar ice. Located atop Mt. Melbourne, Antarctica, the TAROGE-M experiment~\cite{nam2021high} scans downwards for upcoming radio emissions produced by decays of $\tau$ leptons, resulting from near-surface $\nu_\tau$ interactions. Looking up, TAROGE-M also has high sensitivity to down-coming RF emissions associated with air showers. The RET experiment~\cite{Prohira:2019glh}, recently proposed at Taylor Dome, Antarctica, will scan gigaton ice volumes for the radar echo from a neutrino-induced electromagnetic shower evolving in-ice.

Radio emissions resulting from extensive air showers generated by down-coming UHECR provide a `natural' calibration beam for these neutrino experiments~\cite{Barwick:2016mxm}. Even in that case, however, for thermal event trigger rates varying from ${\cal O}$(0.1 Hz) (ARIANNA and RICE, e.g.)$\to{\cal O}$(1--5 Hz) (ARA)$\to{\cal O}$(50--100 Hz) (ANITA) and higher (AURA and SATRA~\cite{landsman2007aura}, e.g.), rejection rates of order
$10^6$--$10^9$:1 are required to ensure that radio signals resulting from UHECR can be identified above background. The remoteness of typical polar experimental sites implies that most anthropogenic backgrounds are mitigated compared with mid-latitude experiments; the fact that human activity is typically confined to a small number of structures further facilitates recognition of such backgrounds. Here we consider whether surface discharges resulting from wind-generated charge separation may present an additional background channel. 

\section{Triboelectric Phenomenology}
Many excellent articles, including experimental measurements of wind-generated electric field magnitudes (EFM), and comprehensive pedagogical introductions to the triboelectric effect can be found in the literature~\cite{lacks2019long,barre1954proprietes,radok1967electrostatic}. Before presenting data from the polar radio-frequency neutrino detectors, we briefly review other experimental tribo-electric field data.

\subsection{Evidence from non-neutrino related field measurements}

Measurements of EFM and electrostatic potentials associated with wind blowing over snow date back to Robert F. Scott's ill-fated Antarctic campaign (1911-1913)~\cite{scott2006journals}. In addition to the quest for the South Pole, that expedition also included extensive scientific measurements, among them long-duration EFM by G. C. Simpson~\cite{simpson1919british} using a near-surface electroscope, which was subsequently correlated with wind velocity {\it ex post facto}. Realizing that the field strength $E_{\mathrm{mag}}$ depended on the wind velocity, Simpson reported the fraction of times $E_{\mathrm{mag}}$ exceeded some threshold value; to ensure that his fractional dynamic range was not saturated, Simpson increased the threshold value, with increasing wind velocity. Simpson did not explicitly report on coronal discharge, although there were contemporaneous studies of such effects by Townsend~\cite{townsend1914xi}.

\begin{figure}
\centerline{\includegraphics[width=0.8\textwidth]{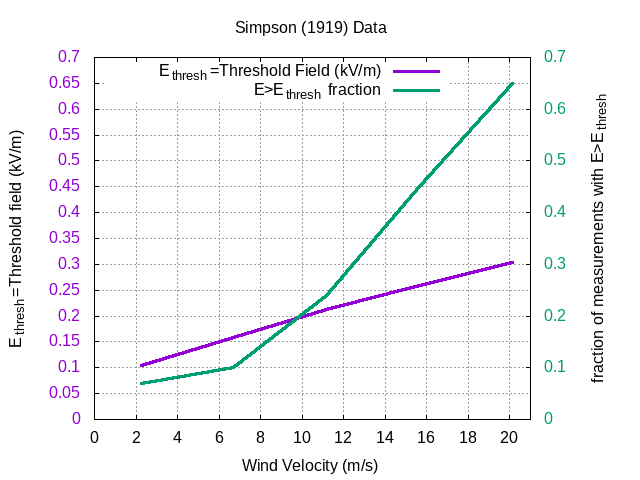}}
\caption{Left scale: Variable electric-field threshold used to reference measurements, as a function of wind velocity (from Simpson \cite{simpson1919british}). Right scale: Fraction of electric field measurements exceeding that threshold. As noted in the original text, the absolute vertical scale was subject to an unknown multiplier, although the relative readings were considered reliable. Note the apparent threshold turn-on, as subsequently observed by other (and with access to significantly more sophisticated instrumentation) experiments.}
\label{fig:simpson}
\end{figure}

Simpson's electroscope data, reproduced in \autoref{fig:simpson}, show the fraction of electric field value measurements exceeding his wind-velocity-dependent voltage threshold. Although the absolute potential readings were uncalibrated, those data indicate a `turn-on' dependence of voltage on wind velocity, with an inflection point at approximately 7--8 m/s, and an electric field direction generally pointing from the ground upwards to the air (${\vec E}=E_{+{\hat z}}$), indicating either a surplus of positive charge on the surface, and/or negative charge in the air above. As noted originally by Simpson \cite{simpson1919british}, this contrasts with the ambient, downward-pointing electric field $E_0$. Zero-wind measurements since then cluster around $E_0\approx$ [-120, -130] V/m at 1~m height, and are attributed to free charges resulting from a combination of cosmic ray-induced atmospheric molecular dissociation as well as radioactive decay ($^{232}$Th, $^{238}$U, $^{40}$K, $^{235}$U) in the Earth's crust. \message{(The 0.03 eV thermal average kinetic energy of an atmospheric molecule is considerably less than the 14.5 eV ionization energy of ice molecules.)}

\begin{figure}
\centering
\includegraphics[width=0.8\textwidth]{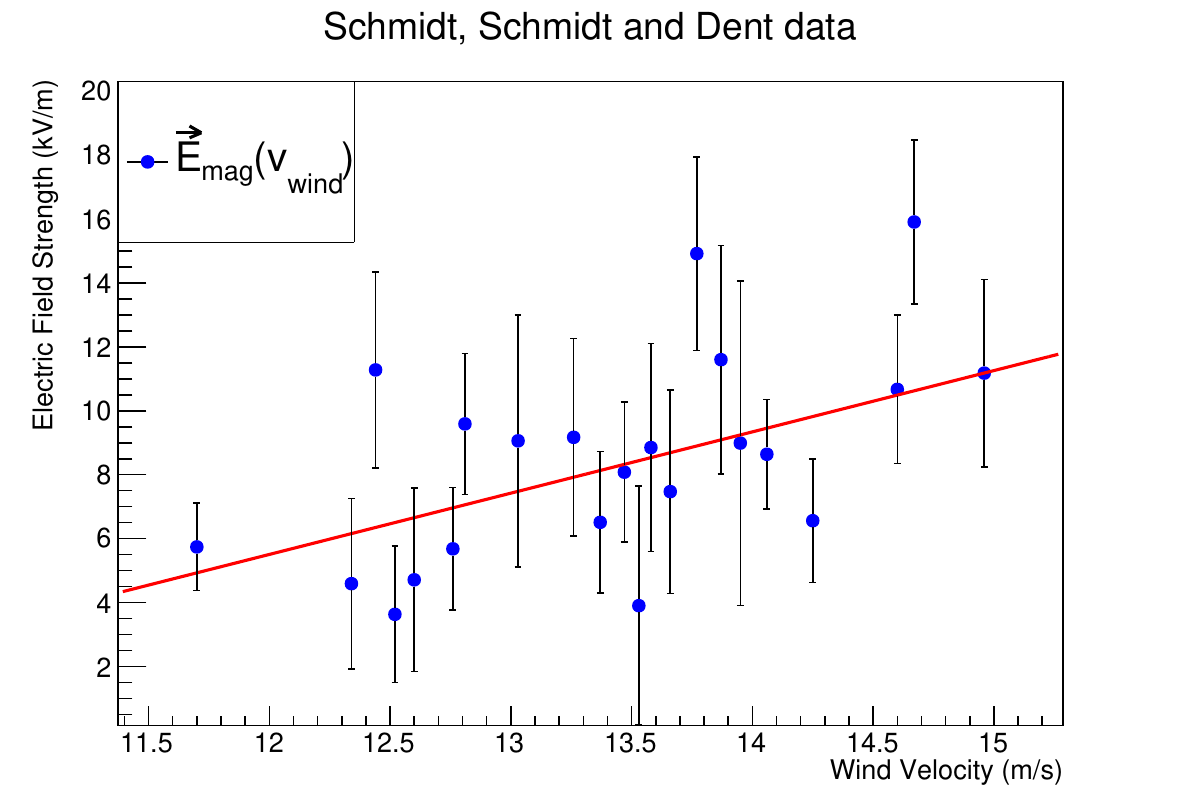}
\caption{Electric Field strength vs.\ wind velocity (data taken from Schmidt, Schmidt and Dent~\cite{schmidt1999electrostatic}). Overlaid is a linear fit (red line) used to determine the threshold wind velocity.}
\label{fig:vwindb}
\end{figure}

Schmidt, Schmidt and Dent~\cite{schmidt1999electrostatic} measured correlations of electric field with both wind velocity and also elevation (50-200 cm) during snow storms. Their published electric field strength vs.\ wind velocity data, for all heights, is reproduced in \autoref{fig:vwindb}. 
To set these data in the context of the field strengths required for atmospheric electric field breakdown, it is useful to compare with the so-called Paschen electric field strength required for coronal arcing through air. At Standard Temperature and Pressure (``STP'', corresponding to 300 K, 1 atmosphere and 0\% humidity [compared with the $<$5\% humidity typical of Antarctica]), this value is about 3400 kV/m at one-meter separation~\cite{paschen1889ueber,townsend1910theory}, considerably larger than the value implied by \autoref{fig:vwindb}. To parametrize this dependence, we consider two models for the threshold characteristics; a) a work function model, in which the kinetic energy of the liberated charges varies linearly with wind speed, vs.\ b) a model in which the kinetic energy of the blowing wind varies linearly with liberated charge, resulting in a quadratic dependence on velocity.
A linear fit of the Schmidt {\it et al.} data yields an x-intercept $v_{\mathrm{wind}}$ threshold value of approximately 9.2 m/s and a slope of 1.92$\pm$0.55 [kV/m]/[m/s].

A similar linear fit has similarly been applied to their electric field vs.\ height data (for all wind velocities in their sample; as evident from \autoref{fig:vwindb}, the authors considered a wind velocity threshold of $v_{\mathrm{wind}}$=11.5 m/s to induce charge separation). The slope of the fit corresponds to a value of -22.9$\pm$7.1 [kV/m]/[m]. These data indicate that the induced potential difference, and presumably the discharge site as well, is likely very close to the surface, and consistent with the 1--10~cm vertical scale of the so-called `saltation' zone. As modeled by Schmidt and Dent, separation and uplift of surface charge requires a minimum threshold mechanical energy transfer\cite{schmidt1993theoretical}).

\begin{figure}
\centerline{\includegraphics[width=0.8\textwidth]{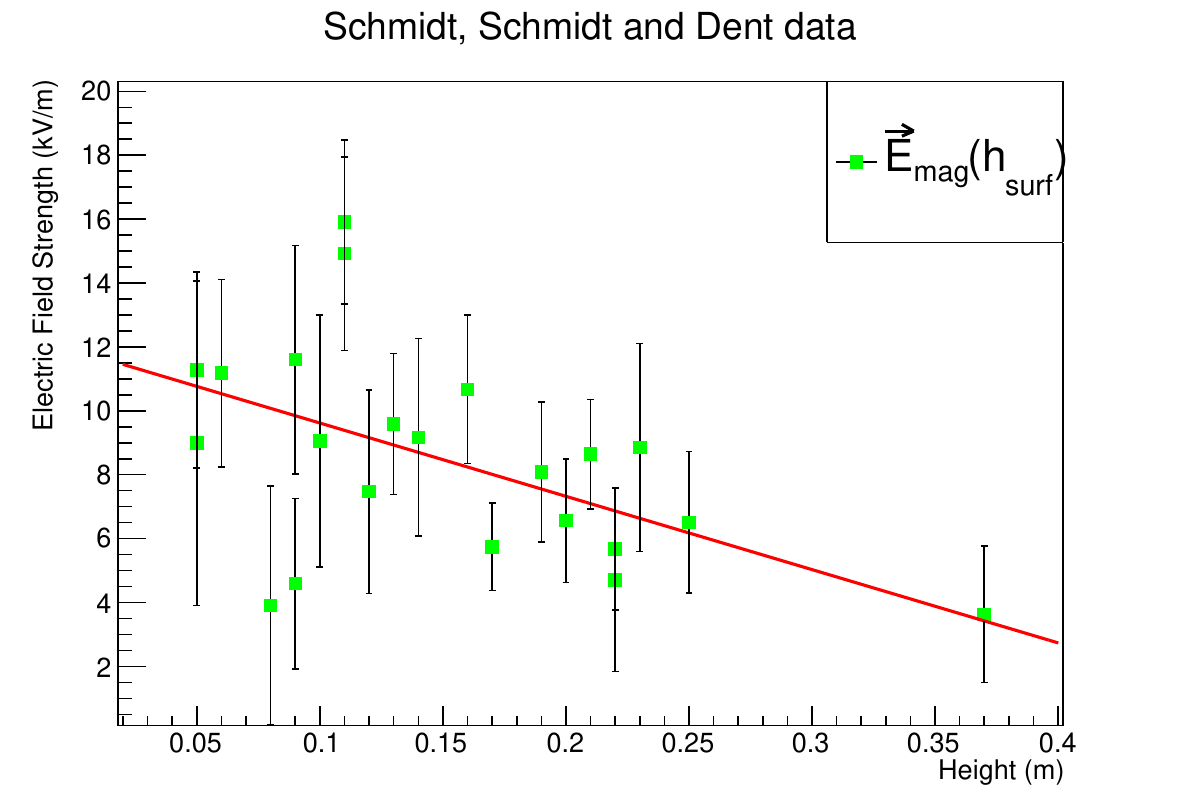}}
\caption{Electric Field strength vs.\ height above surface (data taken from Schmidt, Schmidt and Dent~\cite{schmidt1999electrostatic}). Overlaid is a linear fit (red line) to the data points.}
\label{fig:vwindc}
\end{figure}

Similarly, over a nearly three-month campaign in Canada, Gordon and Taylor~\cite{gordon2009electric} correlated measurements of wind speed and direction, particle flux 10 m above the surface, and the electric field 0.5 m above the surface. Consistent with Simpson's original conclusions, Gordon and Taylor similarly observed a minimum threshold windspeed $\sim$6-8 m/s necessary to produce detectable EFM above ambient, with EFM increasing roughly linearly with wind speed. Gordon and Taylor report a slope $\Delta E/\Delta v_{\mathrm{wind}}\sim$1.5--4.6 [kV/m]/[m/s], and consistent with the Schmidt, Schmidt and Dent data, given the error bars.

Experiments by Latham and Mason (LM, 1961)~\cite{latham1961electric} demonstrated that ice particles blowing over a solid block of ice produce measurable charges on the ice block surface, the ice particles themselves and the air around the ice molecules. LM advanced a thermoelectric explanation, whereby frictional heating of colliding ice molecules leads to dissociation of H$_2$O and re-location of charge, asymmetric between the heavier OH$^{-}$ species and the lighter and more mobile H$^+$, which more readily migrate to colder (surface) regions. 
We note that liberation of both hydrogen atoms from H$_2$0 requires approximately 9 eV, compared with 14--16 eV required for direct liberation of first electrons ("I1") from H$_2$0. 
Absent dissociation, simple ionization of ice (I1$\sim$14.5 eV) conveys electrons upwards from the surface, directly inducing a near-surface gradient. 

\begin{figure}
\centerline{\includegraphics[width=0.7\textwidth]{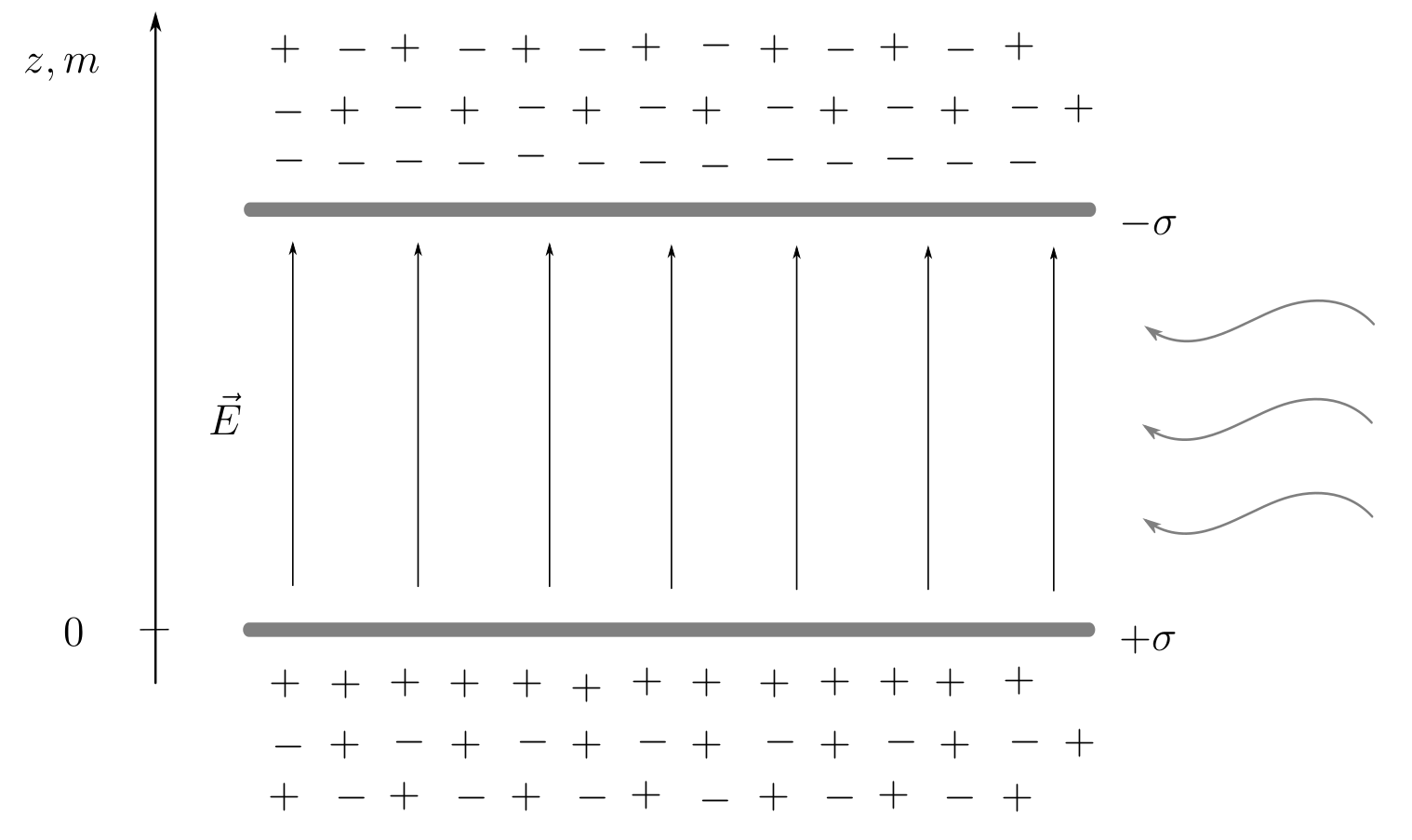}}
\caption{Possible mechanism for charge transport and separation during times of high-winds. Winds displace negative charge upwards, resulting in effective surface-charge densities $\mp\sigma$ above and on the surface, respectively, giving rise to an electric field $\vec{E}$ directed upwards along the z-direction with a typical scale of $\sim$1 m. Not shown is horizontal transport of charge and subsequent precipitation (\emph{saltation}), presumably dependent on wind velocity and direction. }  
\label{fig:cartoon}
\end{figure}

However, simple arguments would seem to disfavor this model -- the thermal velocity of the average air molecule ($\sim$500 m/s, corresponding to a thermal kinetic energy 3/2 kT$\sim$0.03 eV/nitrogen molecule), is significantly smaller than the first ionization energy of ice molecules I1. This can, however, be compensated for, if collisions occur between airborne ice particles rather than individual O$_2$ or N$_2$ molecules. Since a snowflake contains $\sim 10^{18}$ atoms, dissociation of $10^{-7}$ of the atoms in snowflakes would be adequate to produce an electric field gradient of 1 kV/m. \autoref{fig:cartoon} illustrates this simple ionization model.
Competing with direct ionization leading to an electric field vector ${\vec E}=E_{+z}$, 
uplift of H$^+$ and OH$^-$ from the surface should be followed by preferential precipitation of the heavier OH$^-$ ions, favoring ${\vec E}=E_{-z}$.

\begin{figure}
\centerline{\includegraphics[width=0.8\textwidth]{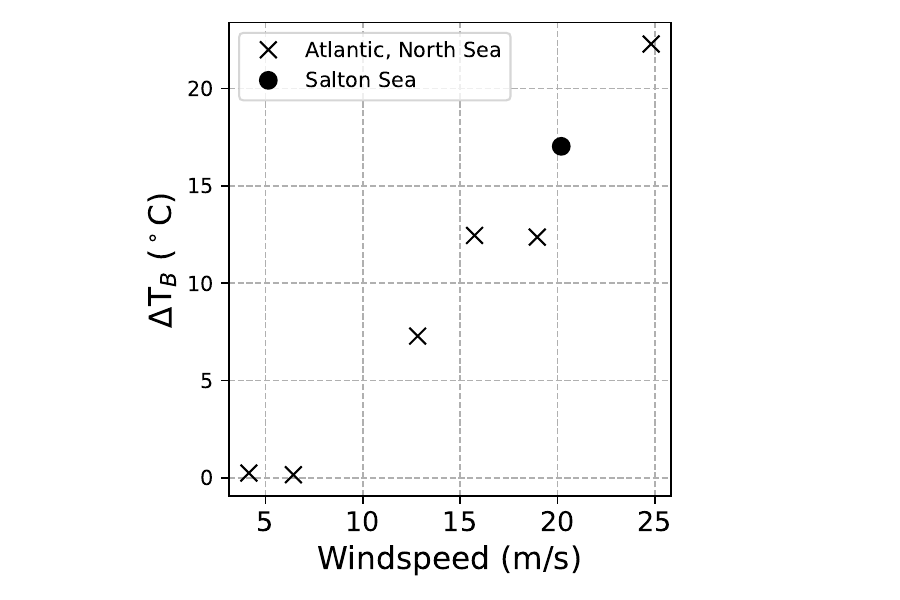}}
\caption{The increase of the brightness temperature (black body equivalent temperature) of the sea surface at 19.35 GHz relative to ambient as function of wind speed (data taken from \cite{radio_noise}).}
\label{fig:vOB}
\end{figure}

Interestingly, wind-velocity threshold effects have been observed in other water-related phenomena. \autoref{fig:vOB} re-produces data illustrating the increase in brightness temperature of the sea surface with wind speed, 
indicating a similar turn-on behavior, both in shape and numerical threshold, and perhaps suggesting a mechanism intrinsic to interactions with H$_2$O molecules. Since enhanced surface brightness typically results from enhanced reflectivity (here, of ambient microwave radiation, and presumably of solar origin), wind-induced charge separation on the sea surface, as outlined above, would similarly enhance the conductivity of the surface, and therefore the observed surface brightness.

\begin{figure}
\centerline{\includegraphics[width=0.8\textwidth]{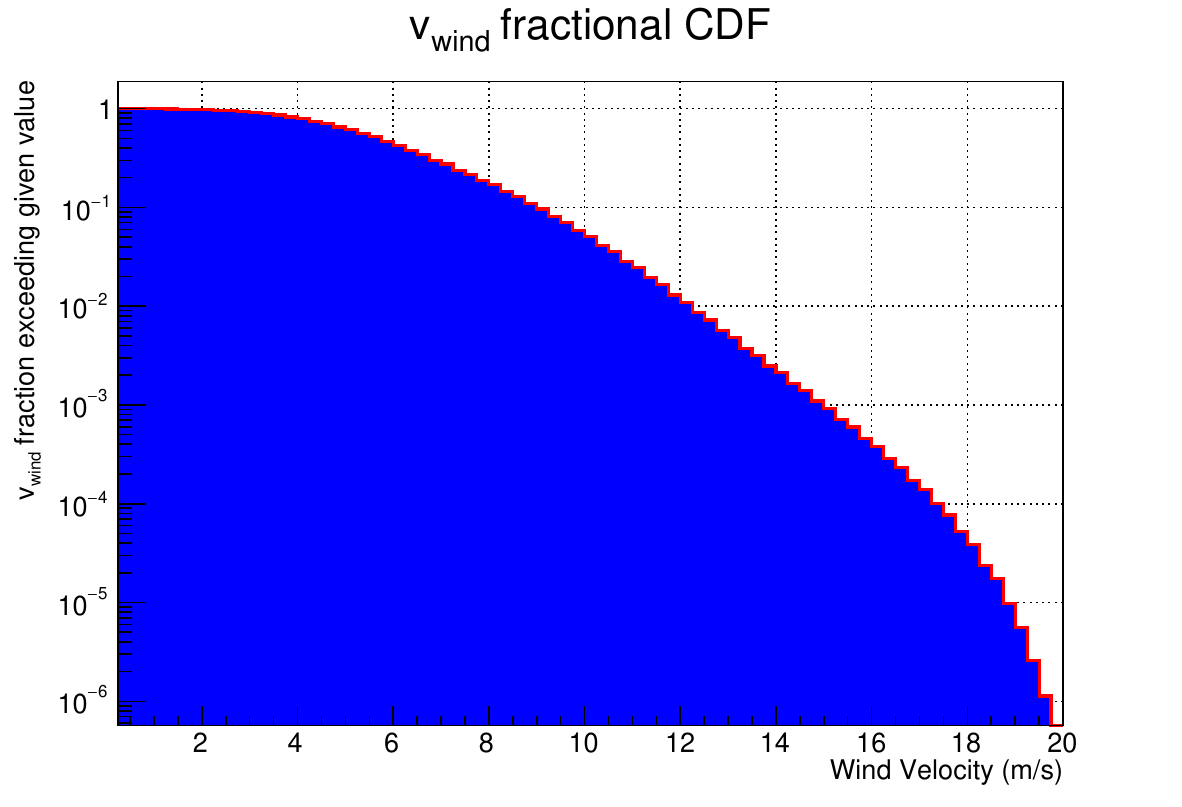}}
\caption{2000-2019 normalized South Pole cumulative wind velocity distribution, showing fraction of times wind velocities exceed a given value. 
Data are taken from \cite{Noaa}.}
\label{fig:SPvwind}
\end{figure}

The bulk of the experimental data above indicates that near-surface electric field strengths increase with wind velocity beyond a `threshold' ($v_{\mathrm{threshold}}\sim$10 m/s). Extrapolation of these results to the polar regions requires consideration of differences in, e.g., mean temperature and humidity -- the average humidity at South Pole is less than 10\% that of the North American northern plains during the winter, for example.
Although the Antarctic interior is typically a low-wind environment (and particularly at the South Pole, as illustrated in \autoref{fig:SPvwind}), there is, nevertheless, considerable high-wind data, 
requiring careful consideration of how radio-frequency experiments might be impacted.

\subsection{Laboratory Tests}
\begin{figure}
\centerline{\includegraphics[width=\textwidth]{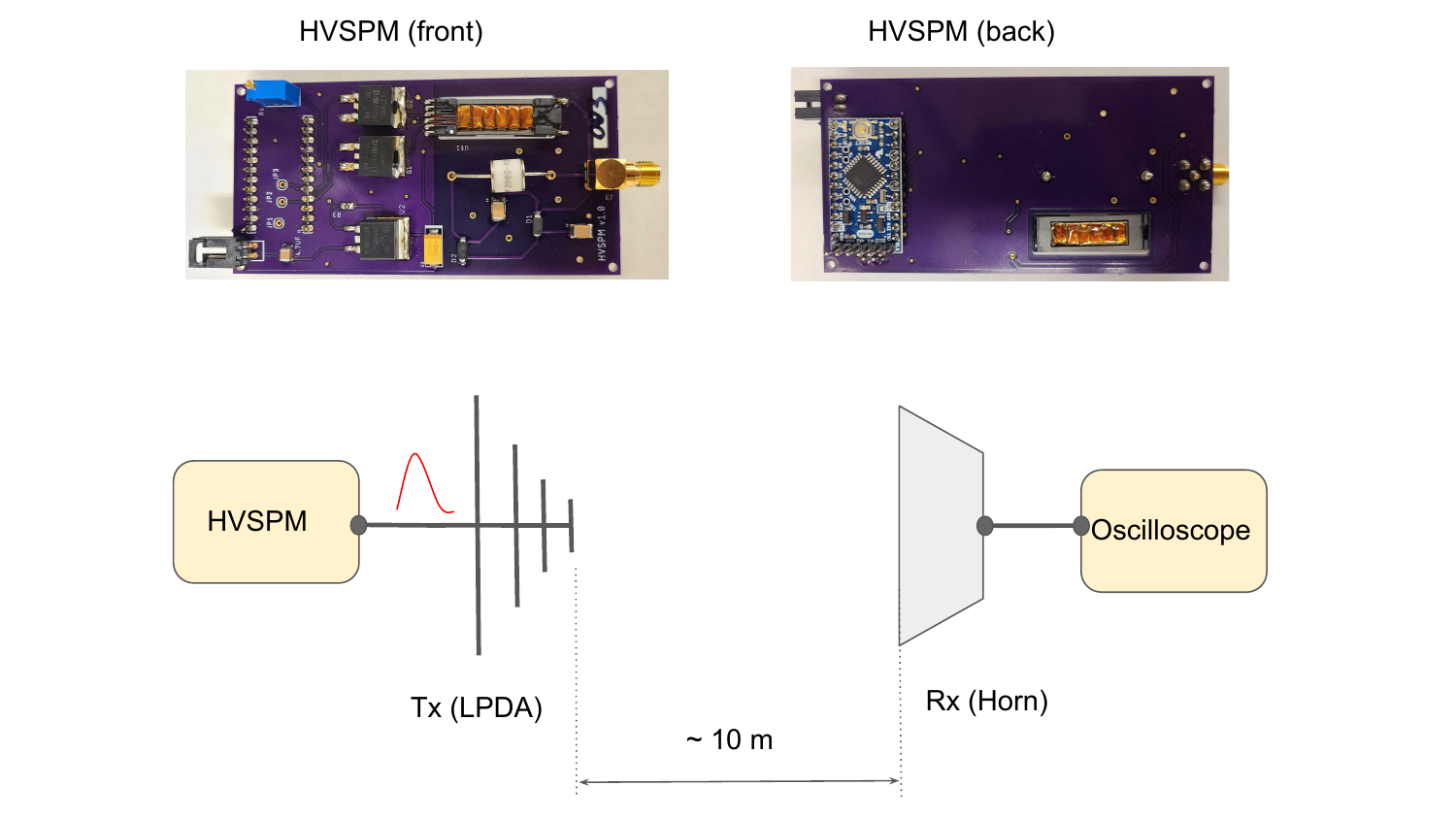}}
\caption{Set-up of laboratory spark discharge tests, showing HVSP-M board and Tx$\to$Rx scale.}
\label{fig:HVSPM-setup}
\end{figure}

\begin{figure}
\centerline{\includegraphics[width=0.8\textwidth]{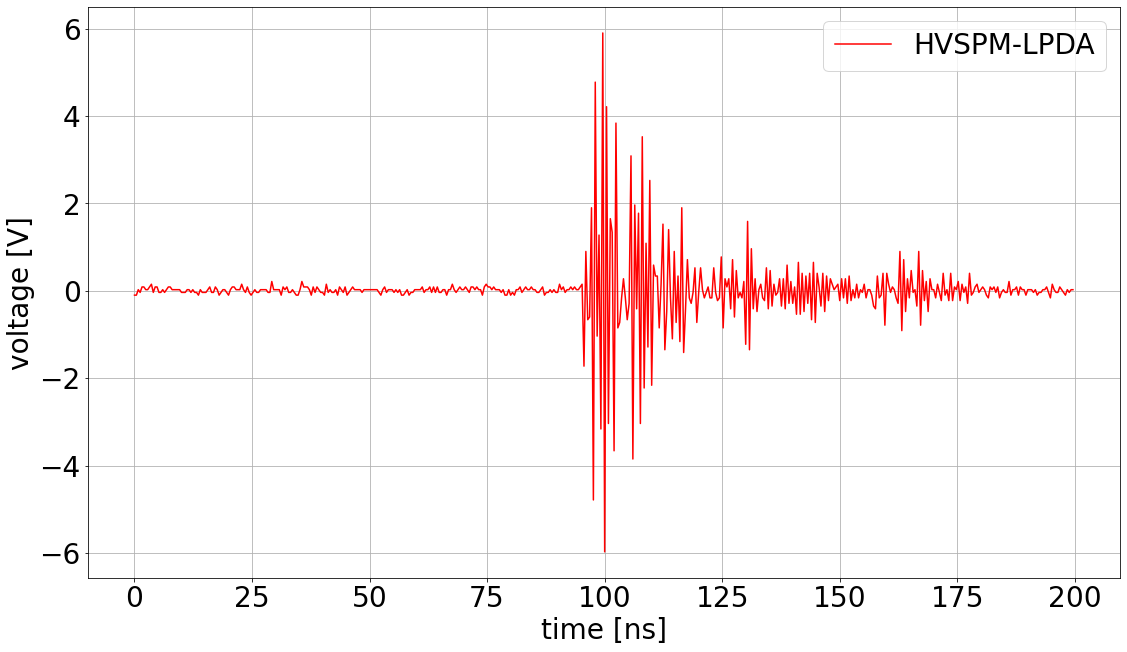}}
\caption{LPDA response (without front-end amplification) to spark discharge generated in the laboratory, using a gas discharge tube (GDT) rated at 800 V nominal signal amplitude. The observed temporal response is that expected for a dispersive LPDA antenna (with group delays of tens of nanoseconds) excited by an impulsive nanosecond-scale signal.}
\label{fig:HVSPM}
\end{figure}

Dedicated laboratory tests were conducted to investigate radio-frequency signals generated by spark discharges in a controlled environment. In the context of experiments seeking measurement of band-limited impulsive radio signals resulting from ice-neutrino interactions, our goal here was to investigate whether such a discharge might generate waveforms with timescales comparable to those expected for signal neutrinos. Gas discharge tubes, such as the Bourns Inc.\ 2095-80-BLF can generate kiloVolt-scale  potential differences at the source, which then undergo rapid discharge within a gas-filled chamber. An 800-V discharge tube was mounted on a custom printed circuit board similar to the High-Voltage SPark (HVSP) board used in previous ice calibration measurements at the South Pole. A custom external 1 pps triggering unit was then constructed to initiate the discharge. The produced radio-frequency signal was then broadcast from a log-periodic dipole antenna (LPDA) identical to those used by both the ARIANNA and RNO-G experiments (see below) to a horn antenna viewing the discharge at a separation distance of 10 meters. \autoref{fig:HVSPM-setup} illustrates the laboratory set-up used to test spark discharge signal characteristics;
\autoref{fig:HVSPM} displays the voltage vs.\ time trace captured by the LPDA, demonstrating $\sim$10-20 ns RF signal time duration, commensurate with the timescale of signals resulting from ultra-high energy cosmic rays and/or neutrinos after convolution with a typical neutrino-experiment transfer function. This test, of course, is only an `existence proof' and does not capture the full range of signal types generated in the field.

\section{Results from radio-frequency neutrino search experiments}
In addition to the aforementioned experiments, understanding the impact of the triboelectric effect on radio neutrino detection experiments is informed by the ANITA~\cite{Smith:2020ecb}, ARIANNA~\cite{Barwick:2014rca}, RICE~\cite{kravchenko2004situ,Kravchenko:2011im,RICE}, AURA~\cite{landsman2007aura}, SATRA~\cite{landsman2007aura}, ARA~\cite{allison2012design}, RNO-G~\cite{Aguilar:2020xnc} and
TAROGE-M~\cite{wang2022taroge} radio detection experiments, as detailed below.

Experimental questions that can be addressed from existing data of these experiments include: 
1) what is the local wind velocity profile at a given site, and how does that correlate with the triboelectric background fraction,
2) do triboelectric backgrounds have characteristics (frequency spectrum, polarization, e.g.) which readily allow identification and rejection 
3) do triboelectric discharges preferentially occur on the surface of man-made structures, or are they distributed randomly, in which case rejection is considerably more difficult, 
4) what are the signal strengths (and, correspondingly, areal scale) observed for triboelectric discharges, and 
5) what is the time lag between the onset of local wind gusts and observable impacts on radio-based detectors?

\subsection{ANITA}
Synoptically scanning the Antarctic continent for upward-coming radio emissions resulting from neutrino interactions in the ice sheet below, the balloon-borne ANITA~\cite{GorhamAllisonBarwick2009} experiment has an instantaneous field-of-view far exceeding that of radio detectors embedded within the ice target itself, albeit with a significantly higher neutrino energy detection threshold owing to the increased average distance-to-interaction (roughly two-orders-of-magnitude greater). Given the isolated, low-noise Antarctic environment, the ANITA high-bandwidth (100 MHz -- 1200 MHz) data acquisition (DAQ) system is designed for high throughput, with triggering very close to the thermal floor, and an architecture achieving $\sim$100 ns deadtimes per trigger. The excellent pointing resolution (approximately 0.5 degrees in both elevation and azimuth, comparable to the in-ice experiments) provides fairly good source localization on the Antarctic continent. A previous statistical analysis of data taken at the time of the ANITA-3 flight indicated an approximately 3$\sigma$ correlation between high wind velocity at a given, reconstructed surface point and the likelihood of ANITA registering a radio-frequency impulsive trigger, although raw waveform data were not reported~\cite{Smith:2020ecb}.

\subsection{ARIANNA}
The simplest radio receiver deployment scheme for neutrino detectors is one in which downward-looking radio frequency antennas are deployed either directly on the Antarctic ice surface, or in shallow, hand-dug trenches ($\sim$1 meter deep).
Using high-gain log-periodic dipole antennas (LPDA) deployed near surface plus a RICE dipole several meters deep, the ARIANNA experiment~\cite{Barwick:2015jca,Barwick:2015ica,Barwick:2014rca,Barwick:2014pca} employs an observation strategy complementary to RICE/AURA/ARA (buried antennas) and ANITA (synoptic viewing). Although the surface receiver strategy results in somewhat compromised effective target volume at high neutrino energies owing to ray optic shadowing effects at near-horizontal incidence angles \cite{barwick2018observation}, this approach offers considerably simplified deployment, ease of antenna retrieval, and freedom in designing high-gain broad-band antennas without geometry restrictions. Subsequent to initial deployment of hardware at Moore's Bay (2009-2015) \cite{Gerhardt:2010js}, two stations were subsequently deployed over a two-year period (2017-2018) at South Pole. Relative to South Pole, wind velocities at the more turbulent Moore's Bay locale exceed 10 m/s approximately 10\% of the time. 
ARIANNA has reported clear evidence for wind-induced triggers in their Moore's Bay receiver array~\cite{Barwick:2014rca,Barwick:2014pca}, characterized by:
\begin{itemize}
\item Wind velocity threshold of 6--8 m/s to produce detectable RF emissions. Above that threshold, mean trigger rates can be enhanced by an order-of-magnitude over the ambient thermal trigger rate, depending on wind velocity.
\item Poor cross-correlation of event waveforms recorded during high-winds with event waveforms expected from neutrino interactions in-ice or UHECR interactions in-air~\cite{Barwick:2016mxm}.
\end{itemize}

Currently, ARIANNA is implementing a microcontroller-based machine-learning algorithm that will provide wind-related background rejection (targeting 1 part per mille) in hardware, at the trigger level, even in high-wind (and therefore high-rate) times. Application of this algorithm to existing data sets has already yielded a promising suppression of wind-triggers, with a similar improvement/reduction in the corresponding event read-out deadtime~\cite{ASTRIDICRC2021}.

\begin{figure}
\centerline{\includegraphics[width=0.7\textwidth]{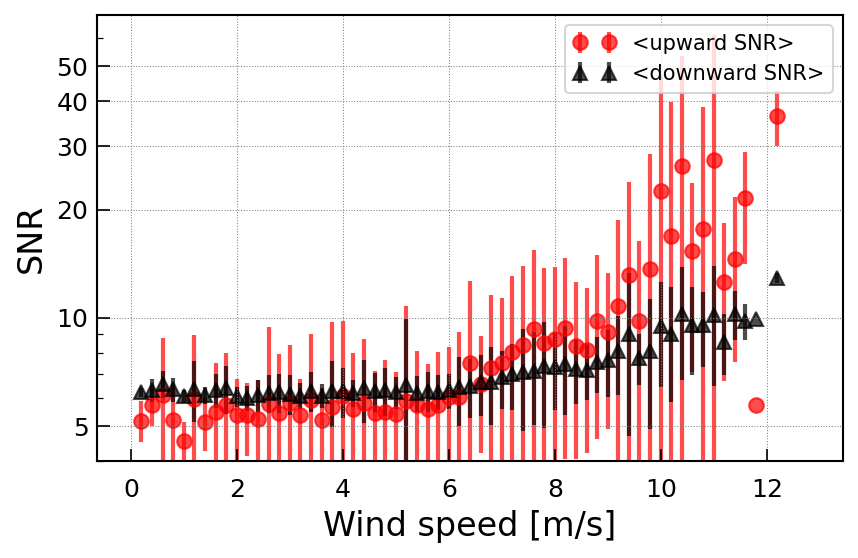}}
\caption{Mean Signal-to-Noise Ratio (SNR) for ARIANNA Station 61 (South Pole) surface LPDA channels, shown separately for upward-pointing vs.\ downward-pointing antennas. As with the RICE and ARA experiments, SNR is defined here as maximum amplitude in a given up- or downward facing channel, divided by the average voltage RMS of the respective channel group in forced triggers. Up- and downward facing channels differ in average RMS, due to the contribution of the Galactic noise. For comparison, on the same SNR scale the radio signal of neutrinos are mostly expected below 10.}
\label{fig:ALPDA}
\end{figure}

 One ARIANNA South Pole station was equipped with both upward-pointing (optimized for radio-wave detection of ultra-high energy cosmic rays) and also downward-pointing (optimized for radio-wave detection of ultra-high energy neutrinos) log-periodic dipole array antennas. The oppositely-directed beam patterns of the two LPDA pairs provide an opportunity to investigate wind-velocity correlations with direction of RF source. The average signal-to-noise ratio (SNR) for the surface LPDA antennas, separately for the upward-pointing vs.\ downward-pointing channels is shown in \autoref{fig:ALPDA}. Similar to RICE and ARA data discussed below, the correlation with wind velocity is most evident for those antennas which are most responsive to on- or near-surface effects. The ARIANNA `turn-on' threshold below 10 m/s may be a direct consequence of an overall enhanced experimental trigger sensitivity for the near-surface ARIANNA antennas to this class of events. Above-surface ARIANNA solar panels are speculated to be a possible discharge site, however, with only two upward facing antennas, the station cannot provide a reliable direction estimate.

In an attempt to rule out interference from a wind-turbine installed at Moore's Bay, it was noticed that the ARIANNA experiment also records triggers from human movement on the snow. For this test, all electronic equipment was turned off at camp (apart from the station itself) and the wind-turbine was rotated by hand, requiring manually paying out a serpentine belt for several tens of meters.
This effect points towards related mechanisms that may cause a rapid electric field change in snow crystals and illustrates the high sensitivity of the neutrino detection systems to such effects.

\subsection{RICE}
The RICE experiment (1995-2011) was the first experiment to search for ultra-high neutrinos using the in-ice radio technique~\cite{Kravchenko:2001id,Kravchenko:2007fy,Kravchenko:2006gf,Kravchenko:2011im,RICE}. RICE primarily comprised 16 `fat-dipole' antennas sensitive over the frequency interval 200-500 MHz at depths of 100-350 meters, over a 200m $\times$ 200m lateral footprint approximately 1 km from the geographic South Pole, and within a radius of 200 meters of the Martin A. Pomerantz Observatory building. Since the RICE antennas were parasitically co-deployed in boreholes drilled for the AMANDA and IceCube experiments, the antenna placement was not optimized for neutrino astronomy. Moreover, the vertical antenna orientation was required to match the form-factor of the holes drilled for the optical experiments, resulting in substantially reduced horizontal polarization sensitivity for the in-ice antennas. 
In-ice signals were conveyed over LMR-600 coaxial cables to a surface data acquisition system based on commercial digitizers, with a typical (deadtime-limited) trigger rate of 0.01-0.1 Hz. Data-taking ceased in December, 2012, when the coaxial signal cables were severed as part of a science upgrade program at South Pole.

\subsubsection{Statistical Measure of Correlations with Wind Velocity}
We numerically assessed the likelihood that observed RICE experimental performance resulted from changes in the wind environment.
We quantify the numerical significance of wind-velocity correlations with experimental observables (specifically, trigger rates and number of antennas with signals exceeding some pre-defined threshold) using a technique similar to an auto-correlation test.

First, a time-ordered array of 
South Polar wind velocities (recorded at 10 minute intervals) $V(t)$ is created. This array is then cross-correlated with a similarly time-sequenced
array of some experimental parameter $P(t)$; the `true' dot product $S_{\mathrm{true}}=\Sigma_i{V_i(t)P_i(t)}$ is 
then compared with the `randomized' dot product
$S_{\mathrm{random}}=\Sigma_{i,j}{V_i(t)P_j(t)}$, for 1000 realizations of the randomized array ${P_j(t)}$, and the fraction of times $S_{\mathrm{true}}$ exceeds $S_{\mathrm{random}}$ tabulated. The significance of $S_{\mathrm{true}}$ can be estimated as the deviation from the mean of $S_{\mathrm{random}}$, given the shape of the $S_{\mathrm{random}}$ distribution.
In principle, several experimental parameters {$P$} might be sensitive to wind velocity, including the instantaneous event trigger rate, the total number of channels of a given polarization with maximum amplitude exceeding some threshold, the root-mean-square voltage in a given channel, the total radio power in a given frequency band, the time that the voltage exceeds some threshold in a given waveform capture, the overall shape of the power spectrum, etc. 

If recorded experimental events are purely thermal in origin, we expect that 
the ${V(t)}$ and ${P(t)}$ distributions should be uncorrelated and $S_{\mathrm{true}}$ should lie within the $S_{\mathrm{random}}$ distribution, modulo complications from episodic anthropogenic backgrounds.  
In such a case, a burst of man-made RF triggers may occur independently of wind velocity, resulting in a false correlation. We also note that the experimental environment at the South Pole underwent signficant changes over the course of RICE data-taking, as various experiments came online, were de-commissioned, or, e.g., station power generators, water retrieval infrastructure, etc, were re-located. The correlations may be reduced by phenomena other than wind, although trends will hopefully remain evident. To ensure veracity of results in our search for correlations of observables with wind velocity, two independent data analyses (``A'' and ``B'') were conducted to cross-check each other. As both analyses yielded similar outcomes, we present results from only one of the two analyses, in what follows below. 

Defining an antenna `hit' as one for which the maximum voltage magnitude measured in a waveform capture exceeds 6$\sigma_V$, with $\sigma_V$ defined as the rms voltage measured from thermal event triggers,
\autoref{tab:RICEXC} displays the trigger rate ($f_{\mathrm{Trigger}}$) correlation results from Analysis A and also
the total number of hit antennas (a.k.a. `receiver hit multiplicity' , or $N_{\mathrm{hitRx}}$) from Analysis B.

\begin{figure}
\centerline{\includegraphics[width=0.8\textwidth]{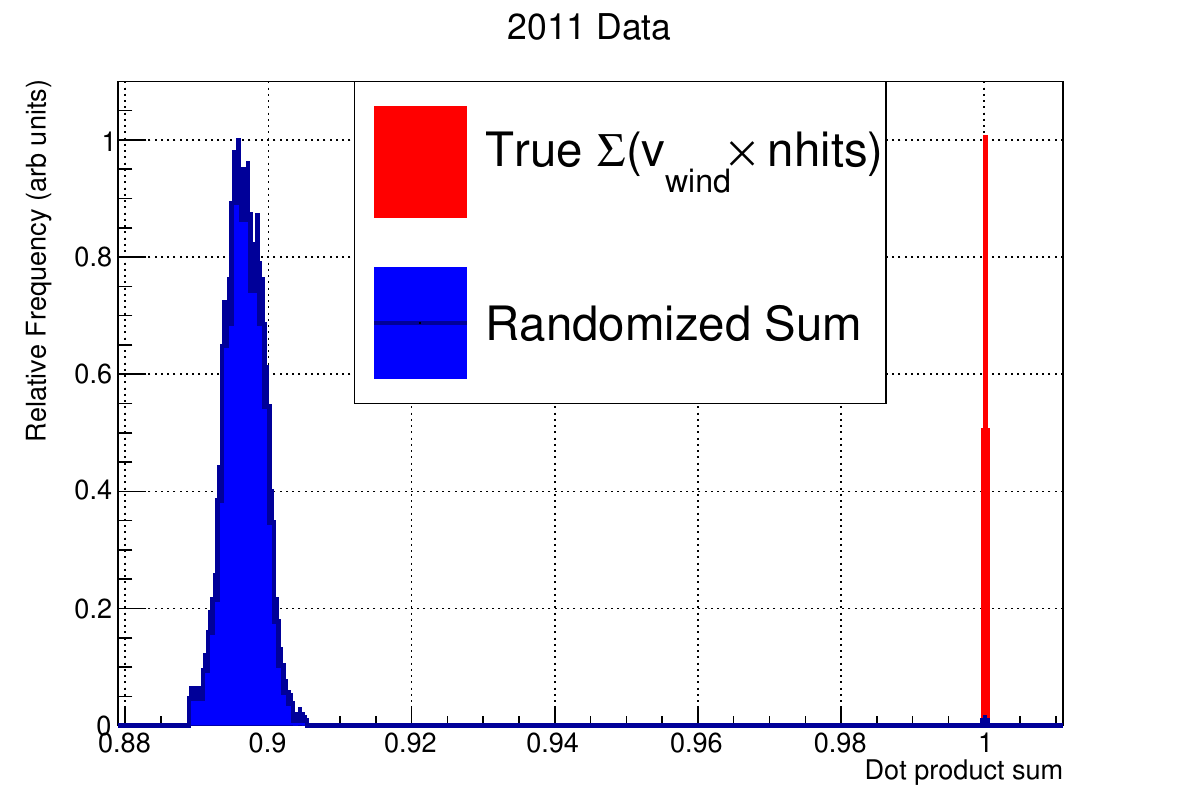} 
}
\caption{RICE experiment product of summed values of wind velocity time profile ($v_{\mathrm{wind}}(\mathrm{time})$) $\times$ average number of hit receiver channels ($N_{\mathrm{hitRx}}(\mathrm{time})$), for randomized time series of $N_{\mathrm{hitRx}}$ (blue histogram) compared with measured series (red), for a typical year (2011). In this Figure, the maximum possible dot produce sum has been normalized to 1 on the x-axis.
\label{fig:cXC}}
\end{figure}

The results of this exercise for one year (2011) are presented graphically in \autoref{fig:cXC}, and show the separation between the `true' value and our `randomized' statistic. All 1000 'time-randomized' dot product sums (blue distribution) are significantly lower than the dot product using the true wind velocity and hit multiplicity time-ordered distributions (red).

Extending this analysis to other years of RICE data-taking, shown in \autoref{tab:RICEXC} is the percentile for `randomized' dot-products $S$ to have magnitude smaller than the true dot-product. A value of 100.0, therefore, indicates that the true dot-product exceeded the randomized dot-product in 1000 test cases, indicating a high degree of correlation with wind velocity. A value of zero, similarly, is statistically unlikely and perhaps indicates the presence of some uncorrelated background which temporally coincides with a low wind-velocity period. Assuming a Gaussian distribution of randomized dot products (consistent with the blue distribution in Figure \ref{fig:cXC}), pure thermal triggers, uncorrelated with wind velocity, are expected to yield values ranging from 5--95.

\begin{table}
\begin{center}
\begin{tabular}{c|c|c}
Year & $f_{\mathrm{Trigger}}$ Percentile & $N_{\mathrm{hitRx}}$ $\chi^2$ Percentile \\
\hline
2003  &  -   & 35.3 \\
2004  & 99.2 & 100.0 \\
2005  & 99.6 & 100.0 \\
2006  & -    & 100.0 \\
2007  & 100.0 & 100.0 \\
2008  & 100.0 & 100.0 \\
2009  & 100.0 & 100.0 \\
2010  & 100.0 & 100.0 \\
2011  & 6.7 & 100.0 \\
\end{tabular}
\caption{Percent of randomized combinations having $S_{\mathrm{random}}<S_{\mathrm{true}}$ for two statistics used to quantify correlation with wind velocity.}\label{tab:RICEXC}
\end{center}
\end{table}


\subsubsection{Source Reconstruction}

RICE uses signal time-of-arrival information on each `hit' receiver antenna to calculate signal arrival directions.
Event reconstruction assumes plane waves incident on the array; typical angular resolutions of 15 arc-minutes in both elevation and azimuth were achieved for englacial calibration pulser signals. For a nearby above-surface source, the in-ice elevation arrival direction depends on the refraction at the air/surface interface, which itself is a function of receiver depth. Our reconstruction algorithms currently do not correct for this receiver-dependent bending. Moreover, the four main buildings closest to the RICE array (the IceCube Laboratory [ICL], the South Pole Telescope [SPT], the main South Pole Station [SPS] building, and the Martin A. Pomerantz Observatory [MAPO], with the last two having very similar azimuthal coordinates) appear as extended sources, smearing the locus of reconstructed directions.

\begin{figure}
\centerline{\includegraphics[width=0.8\textwidth]{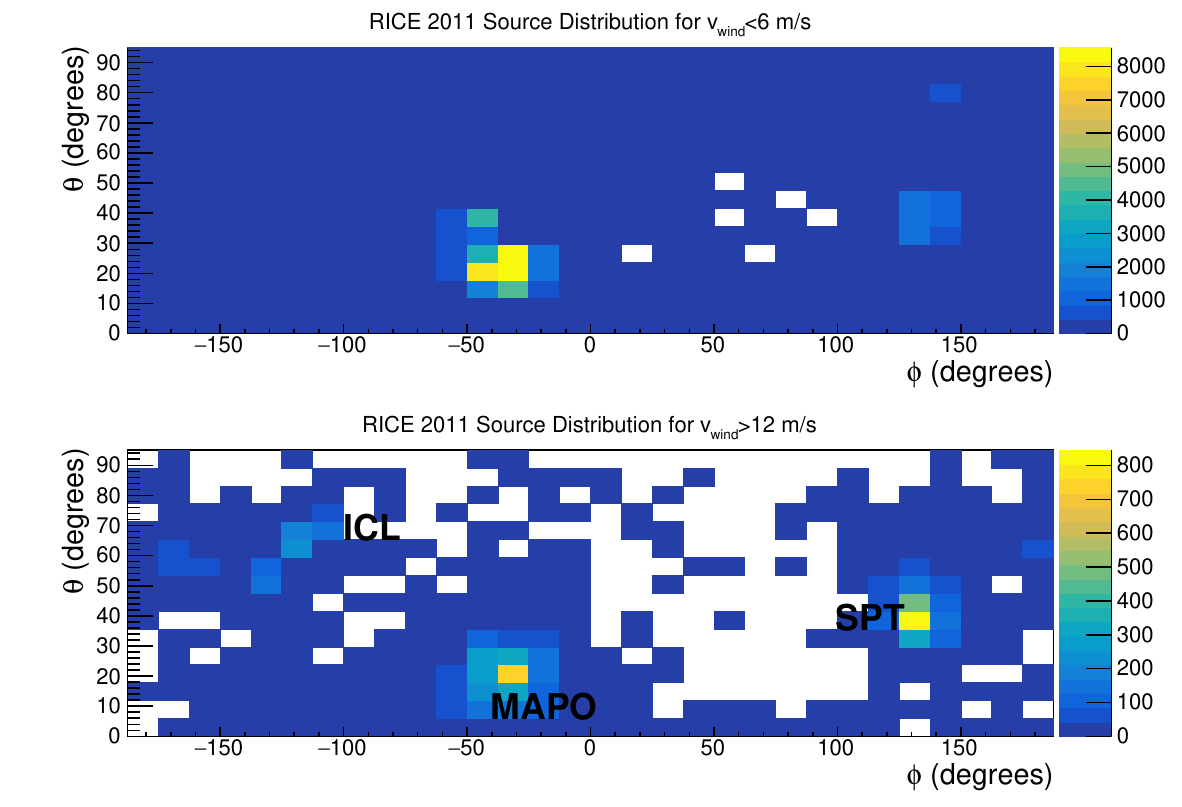}}
\caption{RICE reconstructed source elevation $\theta$ (vertical; degrees) vs.\ azimuth $\phi$ (horizontal; degrees) for 2011 data, comparing event triggers during low-winds (top panel) vs.\ high-wind (bottom panel) conditions. Source distributions are inconsistent with randomly-located surface discharges. At very high winds, we observe more triggers consistent with discharges from the more distant IceCube Laboratory, consistent with higher amplitude discharges during times of higher winds and indicative of the geographical scale over which triboelectric-induced discharges are detectable. The top plot contains 70351 entries vs.\ 8390 for the bottom plot. The ratio of these numbers differs from the fraction of time winds exceed 12 m/s relative to wind velocities less than 6 m/s, since only reconstructable events (requiring four `hit' antennas) enter into these plots.
\label{fig:vwindthephi}}
\end{figure}

Overall, the reconstructed source map (\autoref{fig:vwindthephi}) for high-wind times is inconsistent with the featureless source arrival distribution that would be expected for electrostatic discharges at random locations on the surface. At low wind velocities, the two structures nearest the RICE array (both within 500 meters) dominate the source location map. At high wind velocities, we note that the 0.5-kilometer distant IceCube Laboratory (ICL) is increasingly evident in the source distribution, indicating that triboelectric discharges are detectable over similar km-length scales. 

\subsubsection{RICE Event Characteristics}
\begin{figure}
\centerline{\includegraphics[width=0.8\textwidth]{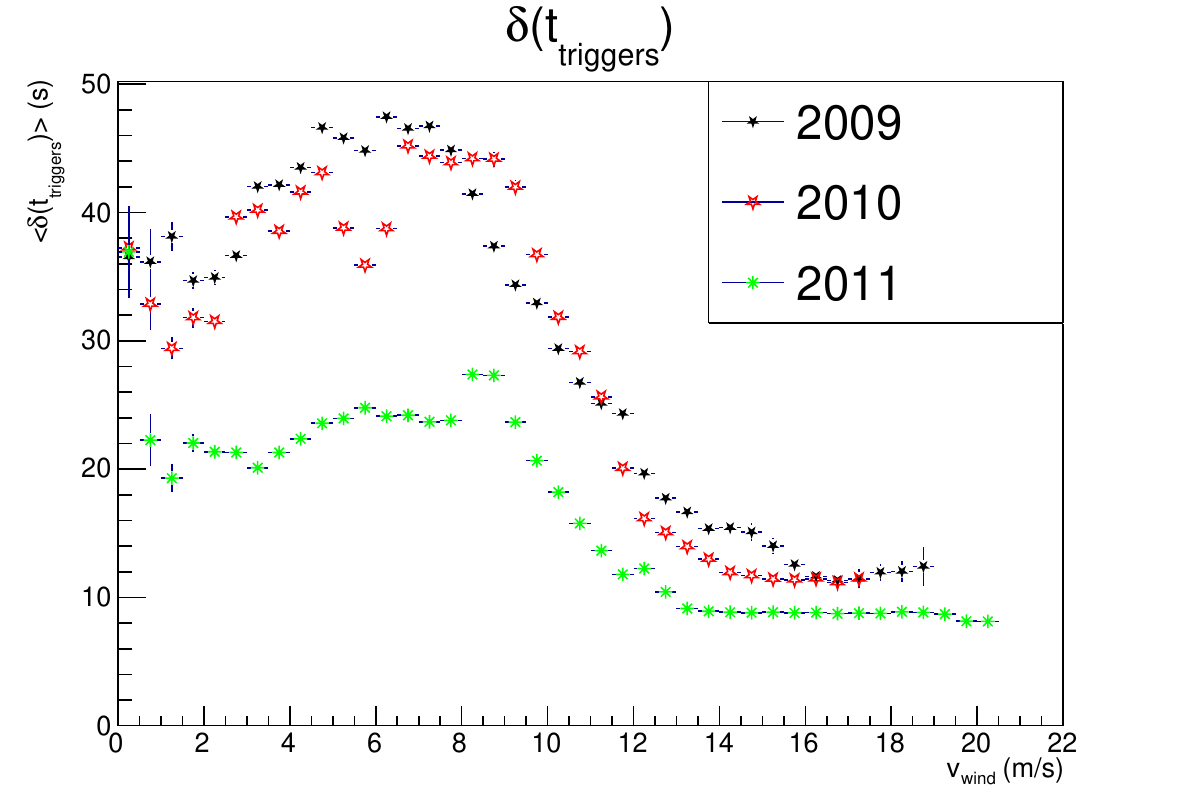}}
\caption{RICE mean time interval between successive triggers (in units of seconds) as a function of local wind velocity (m/s). The minimum time difference (set by the minimum DAQ reset time) is approximately 0.75 seconds per active channel. The absolute trigger rate depends on experimental settings and thus varies from year to year. In RICE data, we note amplitude saturation of the data acquisition system throughput at high wind velocities (as also also noted below for the SATRA experiment).}
\label{fig:windvdttrigs}
\end{figure}

\begin{figure}
\centerline{\includegraphics[width=0.8\textwidth]{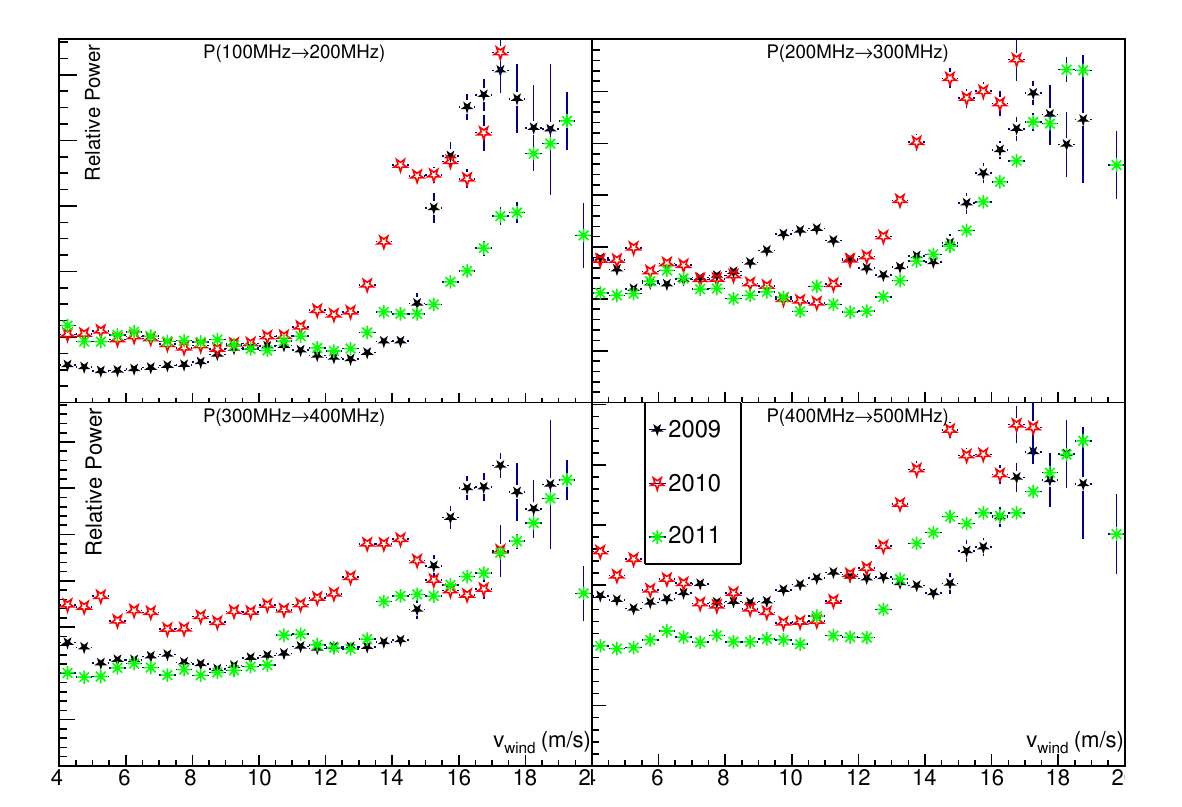}}
\caption{Average RICE waveform signal power, for indicated frequency bins, dependence on local wind velocity. We observe enhanced signal power in all frequency bins, with the largest relative increase at the lowest frequencies to which RICE is sensitive.}
\label{fig:windvPwr}
\end{figure}

We can also characterize the qualitative (and quantitative) effect of wind on recorded event triggers, for the three years of data corresponding to RICE in its most mature form (2009, 2010, and 2011), using a) the time between successive triggers, which has a minimum of approximately 0.75 seconds per active DAQ channel for typical RICE data-taking (\autoref{fig:windvdttrigs}) corresponding to the time interval required for RICE to write data, and b) the fractional power in various frequency bands (\autoref{fig:windvPwr}). At high wind velocities, we observe a significant increase in trigger rate, as well as saturation of the data acquisition throughput. We also observe a trend for events recorded during high wind velocities to have increased power at lower frequencies.

\begin{figure}
\centerline{\includegraphics[width=0.8\textwidth]{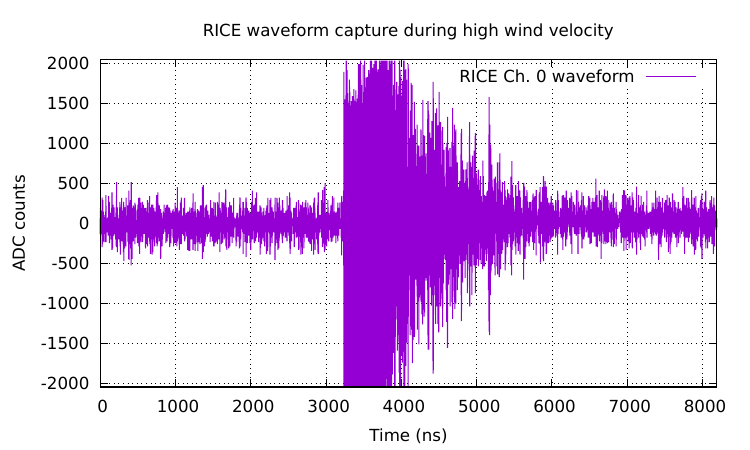}}
\caption{Typical RICE waveform captured during high-winds, illustrating saturation of the dynamic range and extended temporal duration. \label{fig:RICEevwf}}
\end{figure}

\autoref{fig:RICEevwf} displays a waveform captured during a high wind period. Conspicuous in the plot is a signal duration approaching 2 microseconds, or nearly three-orders-of-magnitude longer than the duration expected for true UHEN-generated events. We also observe saturation of the RICE amplitude dynamic range during these times, as well, as evidenced by the `clipping' of the waveform along the y-axis, and suggestive of a very loud, or very local source.

\begin{figure}
\centerline{\includegraphics[width=0.8\textwidth]{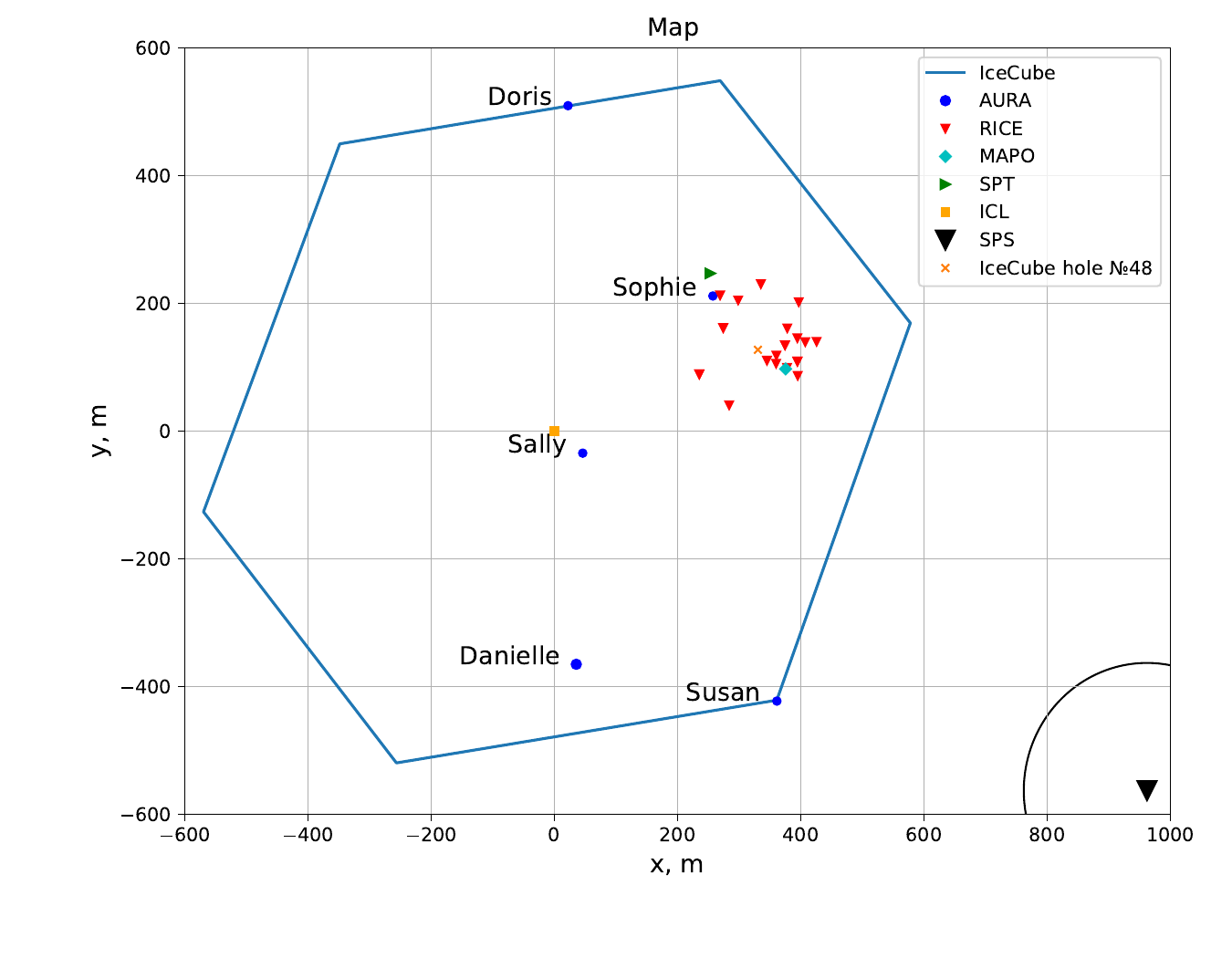}}
\caption{Geometry of AURA receivers relative to IceCube, the South Pole Telescope (SPT), the IceCube Laboratory (ICL), and the Martin A. Pomerantz Observatory (MAPO). Note the extended AURA geographic extent in comparison to the RICE experiment (inverted red triangles).}
\label{fig:geo}
\end{figure}

\subsection{AURA}
The AURA~\cite{landsman2007aura} experiment (2007--2011) receivers were co-deployed in ice holes drilled for IceCube and used a hybrid of RICE (fat-dipole) front-end antennas and IceCube data acquisition electronics, resulting in an increase in the maximum trigger rate by approximately a factor of 1000 relative to the original RICE experiment. Four vertically-aligned antennas, equally spaced over 20 meters, were connected to an IceCube digitizer within an IceCube pressure housing. Each set of four antennas + digitizing electronics comprise a `digital radio module' (``DRM''). Since the antennas are vertically aligned, there is no azimuthal source discrimination, and only elevation angle reconstruction capabilities. Note that use of the IceCube architecture also afforded AURA the excellent time-stamping rendered by the {\tt rapcal} time calibration system~\cite{halzen2010invited}. Five such DRM's (three shallow [``Sally'', ``Sophie'' and ``Susan'', at depths 250-350 m] and two deep [``Doris'' and ``Danielle'', at depth z=--1350 m]) were deployed over 2007-2009; 25\% of the channels failed immediately after deployment. \autoref{fig:geo} illustrates the geometry of AURA, relative to the IceCube experiment.

\begin{figure}
\centerline{\includegraphics[width=0.65\textwidth]{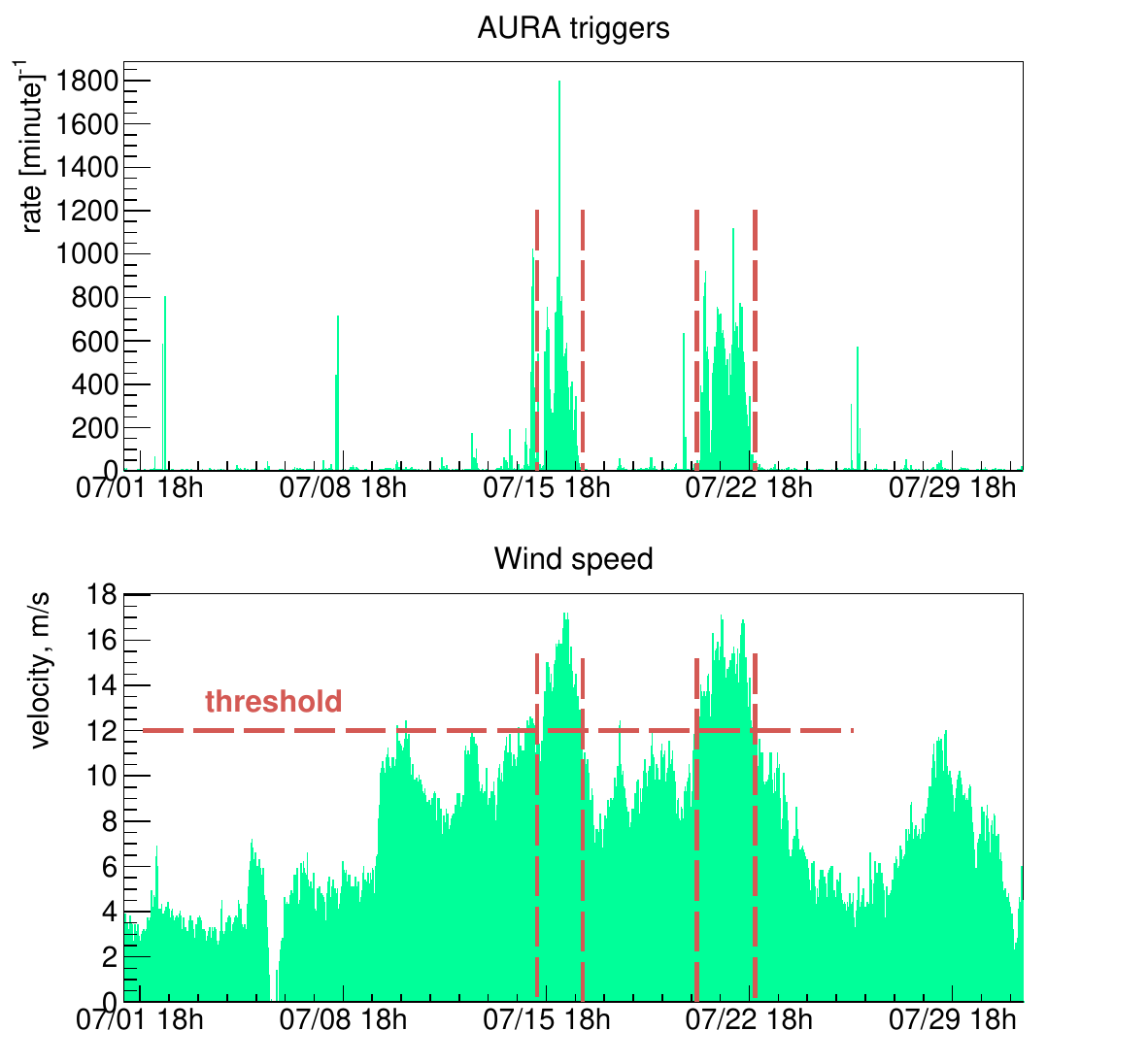}}
\caption{AURA experimental trigger rate, and local wind velocity, as a function of time, for the period from July 1, 2010$\to$July 31, 2010. Increase in wind velocity from July 15, 2010$\to$July 22, 2010 is evident from the plot, as well as associated increase in AURA trigger rates. Also evident in the plot are short-duration spikes in trigger rates, unassociated with wind velocity. Of interest are the periods of extended enhanced winds, with velocities that attain, but do not exceed 12 m/s, which have no evident correlation with trigger rate, suggesting a wind velocity threshold of 12 m/s.}
\label{fig:AURAtrigvWind}
\end{figure}

\begin{figure}
\centerline{\includegraphics[width=0.8\textwidth]{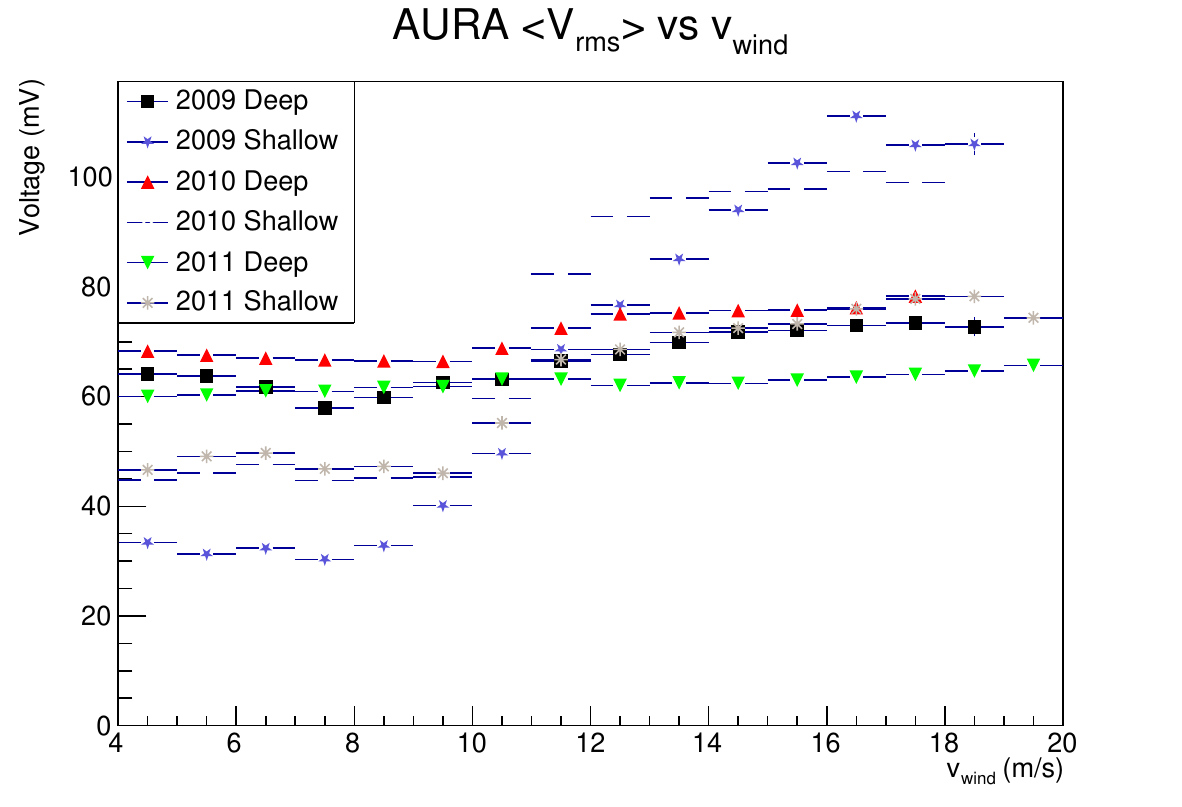}}
\caption{AURA experimental rms voltage as function of wind velocity, for indicated years and radio modules (DRMs). Correlation with wind velocity is most marked for shallow DRMs.}
\label{fig:AURArmsvWind}
\end{figure}

The AURA experiment also showed clear correlations of data characteristics with wind velocity.
\autoref{fig:AURAtrigvWind} shows the average AURA trigger rate as a function of wind velocity; \autoref{fig:AURArmsvWind} displays the 
rms voltage recorded in the (75\%) working DRM channels (there are clearly non-statistical variations in the data points, each of which averages distributions bin-to-bin and are therefore susceptible to broad tails). As before, these distributions are also subject to otherwise uncorrelated episodic anthropogenic noise which may have incidentally flared at some otherwise-random time. Nevertheless,
in both cases, the correlation of wind velocity with both trigger rate and also rms voltage is apparent, and suggest a threshold of 10--12 m/s for observable triboelectric effects for AURA.

We note that, from \autoref{fig:AURAtrigvWind}, within the limits of the time sampling for our wind velocity measurements ($\sim$5 minutes), coronal discharge effects are observed to temporally track wind velocity very closely with no evident hysteresis, or time delay required for charge, or electric field build-up. This observation presumably informs possible models that consider the timescale for build-up of sufficient charge to induce coronal arcing.

\begin{figure}
\centerline{\includegraphics[width=0.8\textwidth]{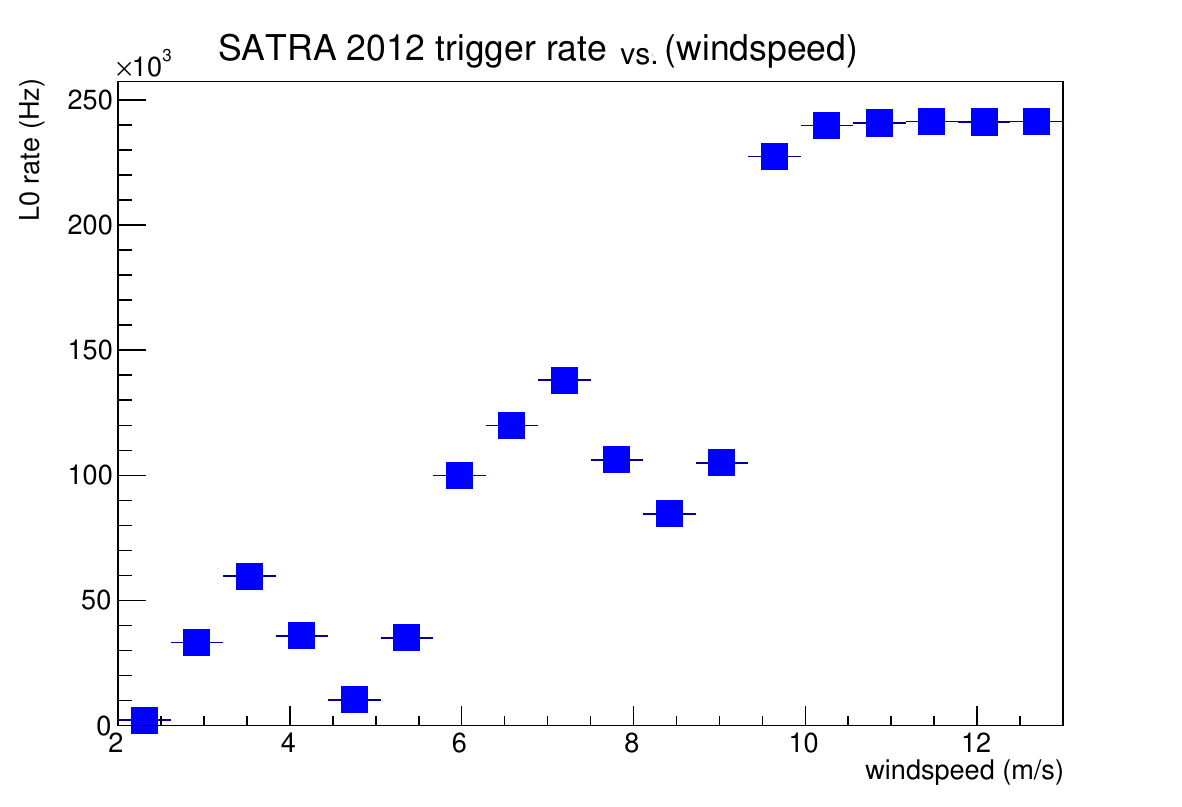}}
\caption{2010 SATRA experimental L0 trigger rate dependence on local wind velocity.The trigger rate saturates above 10 m/s.
\label{fig:SATRAratevWind}}
\end{figure}

\subsection{SATRA}
Like AURA, the SATRA experimental architecture was proferred as a prototype for the next-generation in-ice radio successor to RICE. 
Philosophically, SATRA favored a design with a very dense packing of a large number of sensors,
eschewing long-buffer waveform captures in favor of channel-by-channel measurement of power envelope threshold-crossing times.
This scaled-back data acquisition results in a small event size, and a correspondingly extremely high maximum event-recording rate, which potentially allows these detectors to probe low signal thresholds, and well into the irreducible thermal noise floor.
Unlike AURA, and owing to restrictions associated with IceCube co-deployment, the SATRA antennas were deployed entirely within the firn, at depths not exceeding 50 meters.
The total number of triggers registered every 1.28 seconds is shown in \autoref{fig:SATRAratevWind} for the one year of data (2010) was available for this analysis (although, in principle (and similar to AURA), data transmission capabilities for the in-ice antennas continue to this day). 

Although less definitive, for these near-surface receivers, these data suggest a small contribution of triboelectric events starting at approximately 6-8 m/s. Beyond the threshold of 10 m/s data-taking saturates at the maximum possible rate.

\subsection{ARA}
Initiated with installation of a `testbed' (ARA station A0) in 2011~\cite{allison2012design,Allison:2014kha,allison2016performance,ARA:2019wcf,allison2022low}, the Askaryan Radio Array (2011-) experiment was incrementally upgraded and expanded over the term 2011-2017 to its current 5-station configuration (the testbed was de-commissioned in 2013-2014)
over a 25 km$^2$ footprint at the South Pole. 
Each of the five stations comprises an independent, 16-antenna neutrino detection instrument. Eight ``VPol'' antennas are preferentially sensitive to vertically-polarized signals (i.e., aligned with the ${\hat z}$-axis) and eight ``HPol'' antennas are preferentially sensitive to signal polarizations in the horizontal plane.

Deployed typically to 180-200 m depth, the antennas are sub-grouped into 8 HPol/VPol antenna pairs, with each pair at the vertex of a cube with side length 20 m; the two antennas in a given pair (one HPol and one VPol) are themselves separated vertically by 2--3 m. In contrast to RICE and AURA, which were both deployed within hundreds of meters of the MAPO building, and therefore subject to significant anthropogenic backgrounds, the ARA stations were deployed 2--6 km from both MAPO as well as the main South Pole Station (SPS) and, correspondingly, in a somewhat more benign radio-frequency background environment. 

\begin{figure}
\centerline{\includegraphics[width=0.9\textwidth]{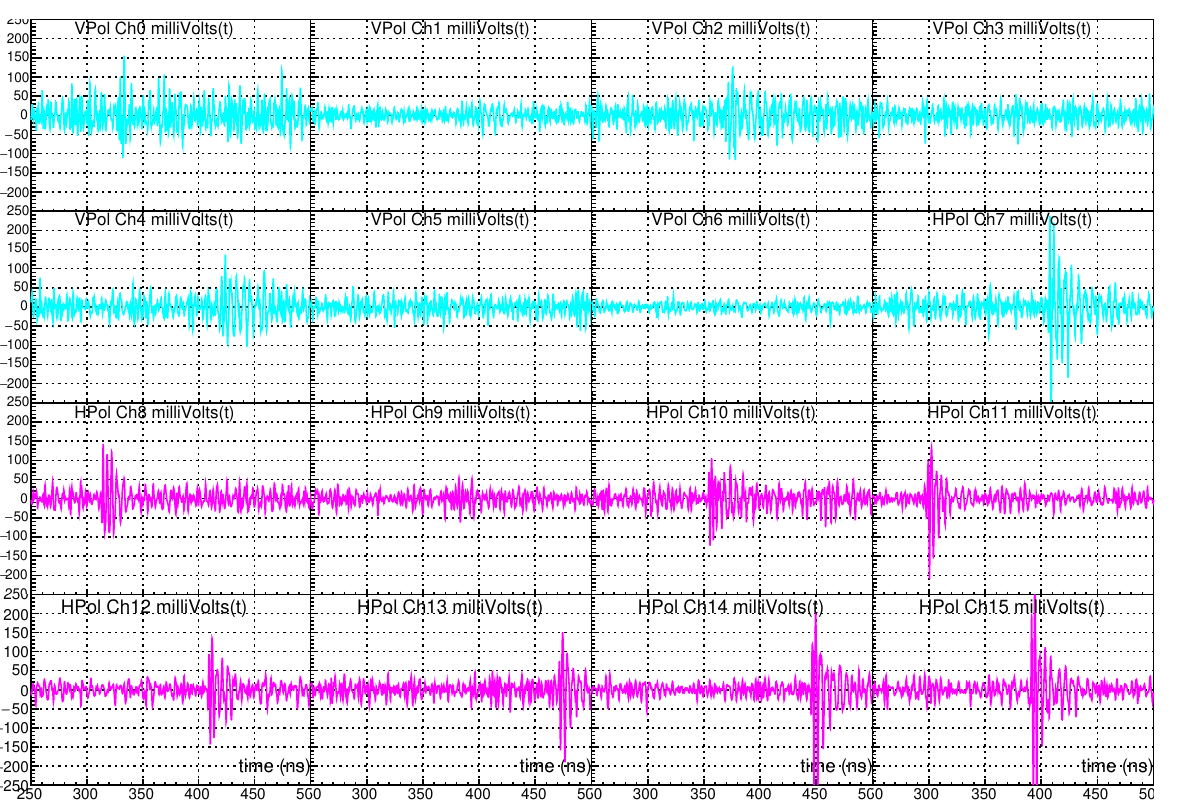}
}
\caption{16 channel ARA event display for event trigger a) recorded during high winds and b) having particularly impulsive time-domain signals. This event exhibits significant power in both VPol (top two rows) as well as HPol (bottom two rows) channels, and reconstructed to the location of Wind Turbine 3, which is the nearest metal structure. During these data-taking, that wind turbine was not connected to the power grid, but a dummy load used to shed turbine power. Originally found in an ARA UHECR search, this event was later re-classified as likely background in origin.\label{fig:alisa}}
\end{figure}

Triboelectric event background candidates have previously been reported from ARA data~\cite{latif2020towards}. A dedicated search for radio emissions from down-coming cosmic-rays interacting in the atmosphere reported 11 radio signals having characteristics consistent with those expected for down-coming geomagnetic radiation from UHECR \cite{latif2020towards}. 
\autoref{fig:alisa} displays the 16 channels of waveform information for a UHECR candidate captured by ARA station A3 on a day (August 25, 2014) when wind velocities exceeded 15 m/s. In the display, the top two rows correspond to the voltages recorded by the vertically polarized antennas, as a function of time, whereas the bottom two rows correspond to the voltages recorded by the horizontally polarized antennas. Since the top/bottom row in each polarization pair corresponds to the upper/lower layer of four deployed antennas (with the two layers vertically separated by 20 m, as detailed above), the leading edges of the observed signals clearly correspond to a down-coming signal, with signal onsets in rows 1 and 3 leading the signal onsets in rows 2 and 4, respectively.

Conspicuous in this event is the presence of signal power, above background, in both the vertical (top two rows, cyan), as well as the horizontal (bottom two rows, magenta) polarizations. We also note relatively sharp risetimes; such short time scales are atypical of high-wind event triggers. Three of the UHECR candidates were eventually discarded as likely wind-related events, as they triggered at times of high winds (greater than 12 m/s).
All three of the candidate events reconstruct to the location of a wind turbine erected at South Pole in 2011 to investigate the potential for renewable power provision through the Antarctic winter~\cite{Besson:2014gda}. The turbine was inactive at the time these event triggers were recorded, suggesting that the observed signals may have had their origin in discharge on the surface of the metal wind turbine tower itself.
Quantitatively, given the wind velocity distribution shown in \autoref{fig:SPvwind}, we can calculate the probability that three events out of 11 total UHECR candidates would all occur when wind velocities exceeded a threshold of 12 m/s (chosen based on prior observations from the RICE and AURA experiments). 
The likelihood of k events, each with a probability P, being observed from a total sample of n events is well-known from standard probability: P(k)=[n!/(k!(n-k)!)]${\rm P}^k({\rm 1-P})^{n-k}$. In the UHECR search cited above, n=11, k=3, and, averaged over the years for which this analysis was conducted, P=0.026, yielding P(2)=0.3\%, which is a measure of the probability that there is no triboelectric contamination of our UHECR candidate sample.

We have used ARA testbed data to further investigate the polarization content of triboelectric-generated RF emissions, as well as the signal strength proximal to the surface. We group the deployed testbed antennas, as follows: 
\begin{itemize}
    \item[a)] bicone antennas aligned along the borehole axis deployed to depths of 25--30 meters below the surface, and preferentially sensitive to vertically polarized electric fields (``Deep VPol'')
    \item[b)] Quad-slot cylinder antennas co-deployed in the same boreholes as the bicones at similar depths, but preferentially sensitive to horizontally polarized electric fields (``Deep HPol'')
    \item[c)] Quad-slot cylinders deployed in separate boreholes (``Deep Hqsc''); in contrast to the ``Deep VPol'' and ``Deep HPol'' data acquisition channels, which have a 150--1000 MHz passband, these channels have a 100--450 MHz passband and do not participate in trigger formation
    \item[d)]`fat' dipole antennas, based on the original RICE design, and laying horizontally in shallow trenches close to the snow surface (``Surf HPol''), with a 25--300 MHz passband.
\end{itemize}

\begin{figure}
\centerline{\includegraphics[width=0.8\textwidth]{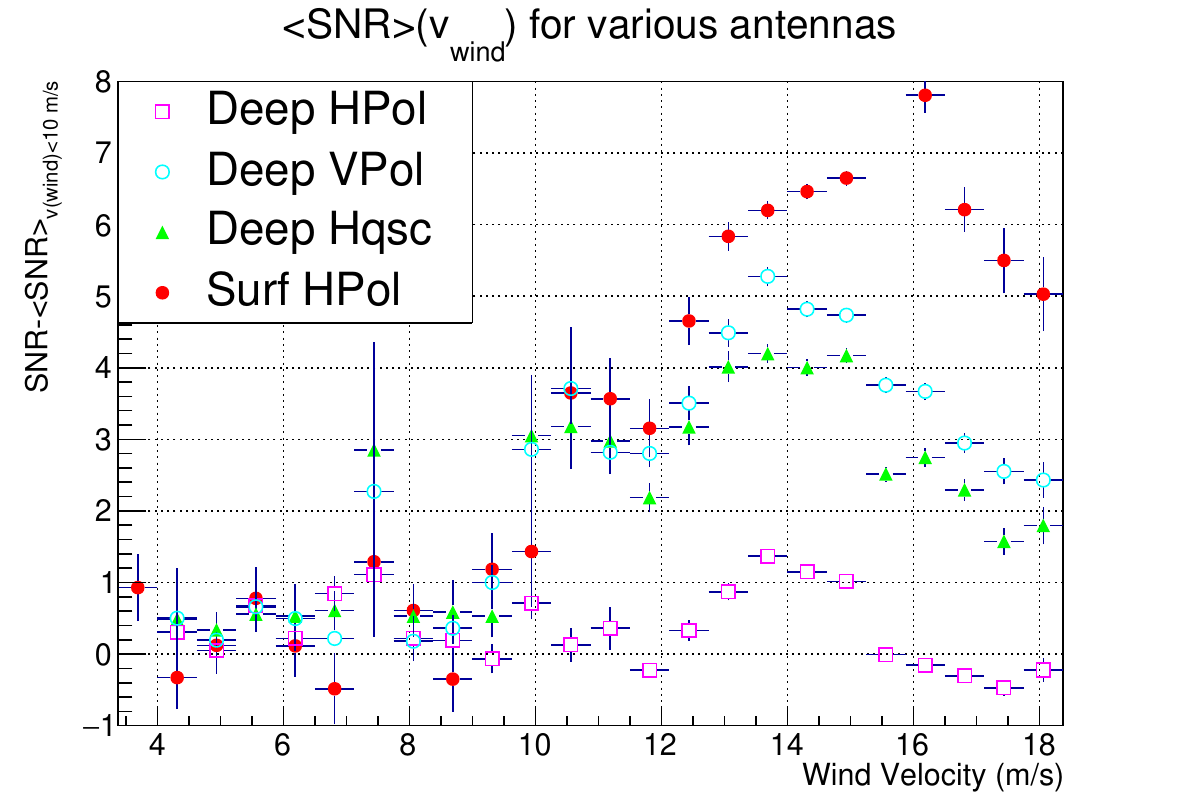}}
\caption{Background-subtracted signal-to-noise ratio (here, defined as the magnitude of the maximum voltage excursion divided by the rms voltage in a given channel, minus the mean SNR as measured in thermal noise events), as a function of wind velocity, for various ARA testbed antennas, as detailed in the text. 
\label{fig:A0911}}
\end{figure}

As shown in \autoref{fig:A0911}, the signal-to-noise ratio shows a clear enhancement for wind velocities exceeding $\sim$10 m/s. Consistent with other observations that emissions associated with triboelectric discharges preferentially favor lower radio frequencies, the effect is most noticeable for those antennas with lower-frequency acceptance. We note that the antenna frequency response for the Deep VPol bicones is approximately twice as broad as the frequency response of the Deep HPol antennas, approximately consistent with observation. Interestingly, both the ARA and RICE data (Fig. \ref{fig:windvPwr}) are suggestive of a peak correlation effect at $v_{\mathrm{wind}}\approx$15-16 m/s; at higher wind velocities, the correlation in the normalized ARA SNR distribution drops, for example. We speculate that at higher wind velocities, surface charges are more geographically dispersed relative to their formation site.

\begin{figure}
\centerline{\includegraphics[width=0.8\textwidth]{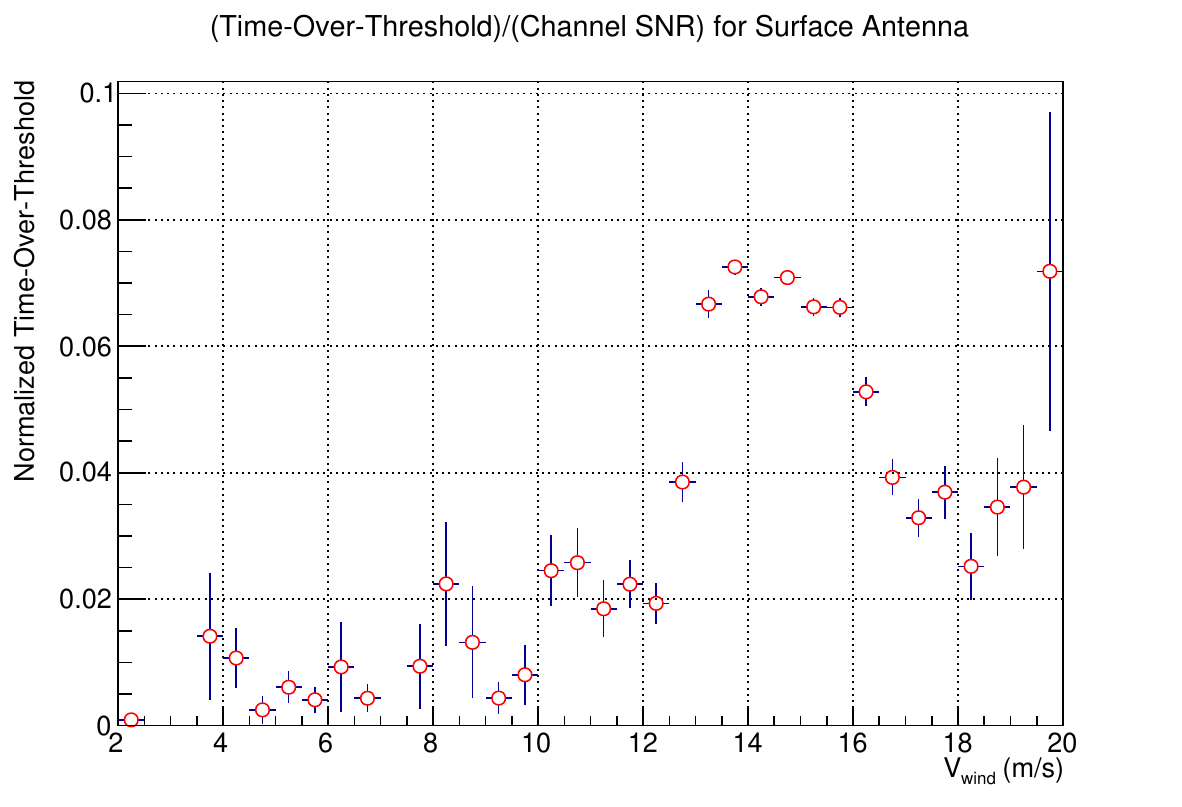}}
\caption{Waveform time-over-threshold, normalized by signal-to-noise ratio for a surface channel of ARA, as described in text.}
\label{fig:windvfTOT}
\end{figure}

As noted already by ARIANNA, signals registered during high wind periods are typically considerably longer than the 1--10 ns timescale characteristic of either UHEN- or UHECR-induced signals. We quantify this using the ratio $f_{\rm TOT}$/SNR, with $f_{\rm TOT}$ defined as the fraction of samples in a given waveform with voltage exceeding six times the rms-voltage $V_{\mathrm{rms}}$ in that waveform and SNR the ratio of the maximum voltage magnitude excursion divided by the rms noise in that channel. For a UHECR- or UHEN-induced signal, typical values for $f_{\rm TOT}$ and SNR are 0.04 and 10, respectively, such that the ratio is typically approximately 0.004. Large amplitude excursions will increase $f_{\rm TOT}$, but, if the signal shape is unchanged, SNR will also increase, such that the ratio should be relatively constant. As shown in \autoref{fig:windvfTOT}, we observe extended waveforms, without a compensating increase in the signal-to-noise ratio, and consistent with the RICE observations of waveform shape.

\begin{figure}
\centerline{\includegraphics[width=0.9\textwidth]{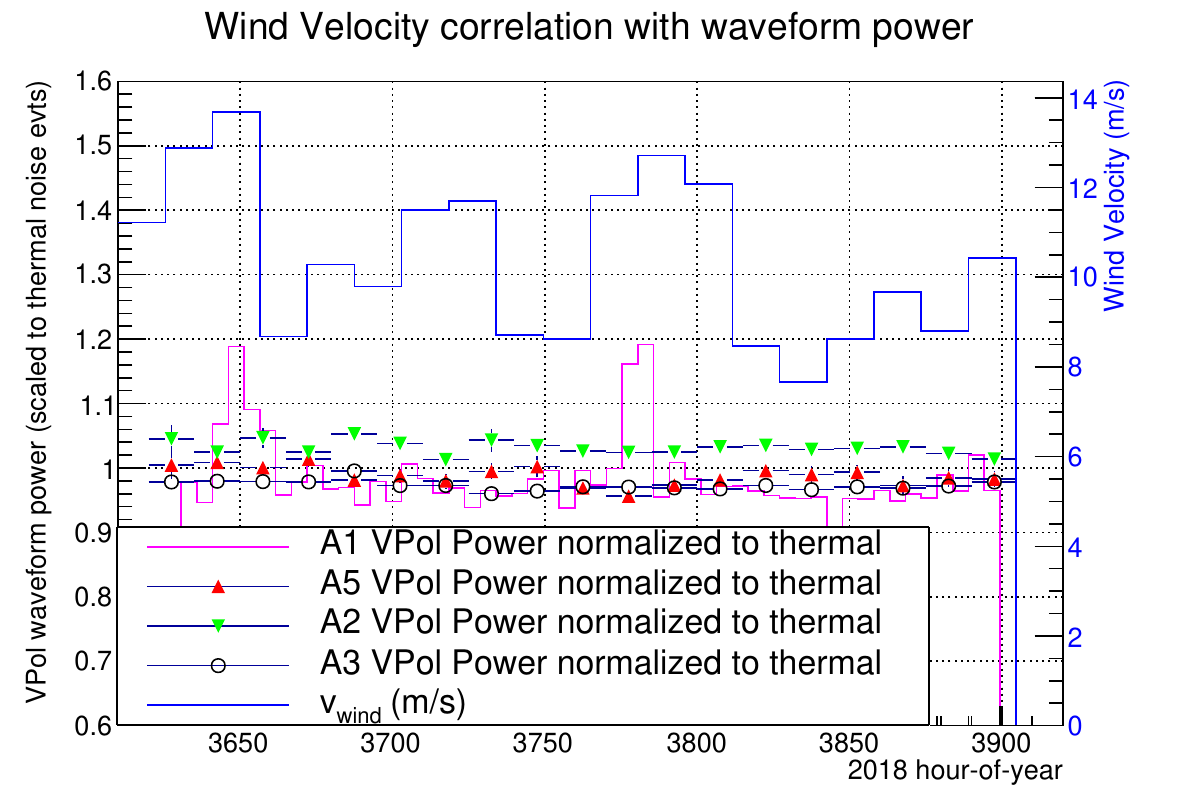}}\caption{June 2018 correlation between ARA stations' waveform power for four-hit events and registered south polar wind velocity. The only station that indicates a correlated enhancement in signal amplitude with high wind velocities is ARA station A1, which is also closest to South Pole Station. Trigger rates for all remaining ARA stations ($>$2 km distant from SPS) registered no correlation with wind velocity.
\label{fig:A1A5June2018}}
\end{figure}

Of particular interest to the planned radio array of IceCube Gen-2 is the question of whether triboelectric discharges can be mitigated by siting experimental hardware away from above-surface conducting structures. Data taken in June, 2018, when all five ARA stations (A1-A5) were active, and wind velocities exceeded 14 m/s on several occasions, is informative in this regard. As the ARA stations were successively constructed increasingly remotely relative to South Pole buildings, if discharges occur preferentially on man-made structures, we would expect station A1 to have the most obvious correlation of event characteristics with wind velocity. \autoref{fig:A1A5June2018} shows the wind velocity distribution overlaid with the total waveform power ($\Sigma(V_i^2)$ for all the voltage samples in a waveform) for the vertically-polarized antennas, also as a function of time.
Wind velocity-correlated signal is only evident in A1, located within 1 km of Wind Turbine 3. It is not observed in the other, more distant ARA stations.

\subsection{Coincident Events between RICE, AURA, and ARA}
\begin{figure}
\centering
\includegraphics[width=0.9\textwidth]{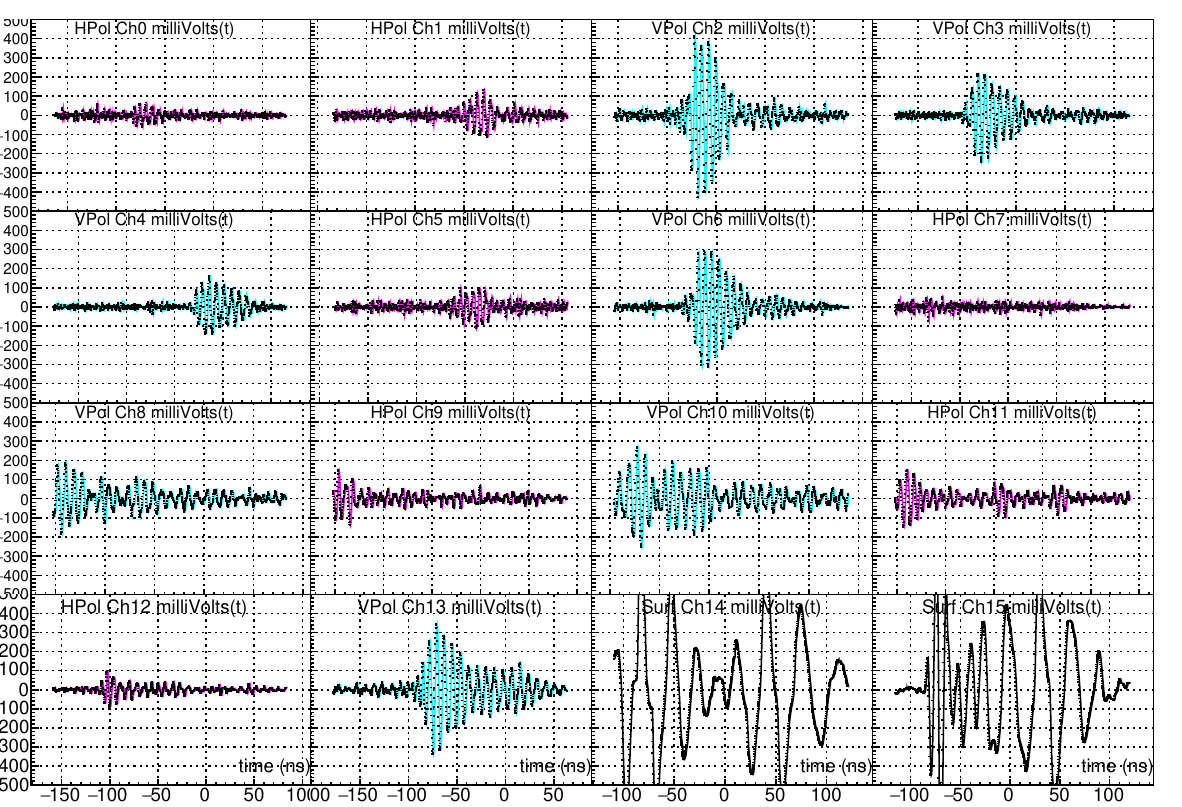}
\caption{ARA testbed high wind event, recorded Sept. 21, 2011, 12:40 UTC. The wind velocity at the time this event was recorded was approximately 13 m/s. Note the large signals recorded in the surface antenna channels 14 and 15 (lower right black traces), which have frequency response extending down to 25 MHz. For other channels, signal is generally more prominent in VPol channels (cyan) vs. HPol channels (magenta).}
\label{fig:ARA21092011evt}
\end{figure}

Radio-frequency emissions registered coincidentally by multiple experiments ensure that triggers are not due to noise generated locally within a single experiment's data acquisition system (DAQ). During mid-September 2011, when South Polar wind velocities reached 17 m/s, RICE, AURA and ARA testbed were all active, although the ARA testbed DAQ clock was not synchronized to UTC global time. \autoref{fig:ARA21092011evt} displays a typical testbed waveform recorded during that period, showing significant signal power in both the VPol as well as the HPol channels.

Among the characteristics distinguishing true cosmic ray signals from tribo-electric background events is the relative signal power as a function of polarization. Whereas signals from neutrinos and ultra-high energy cosmic rays are preferentially vertically or horizontally polarized, respectively, our studies indicate that tribo-electric emissions have a more uniform polarization distribution. The September, 2011 wind storm, for example, fired (separately) both the triggers sensitive to either vertical polarization vs. horizontal polarization. 

\begin{figure}
\centerline{\includegraphics[width=1.0\textwidth]{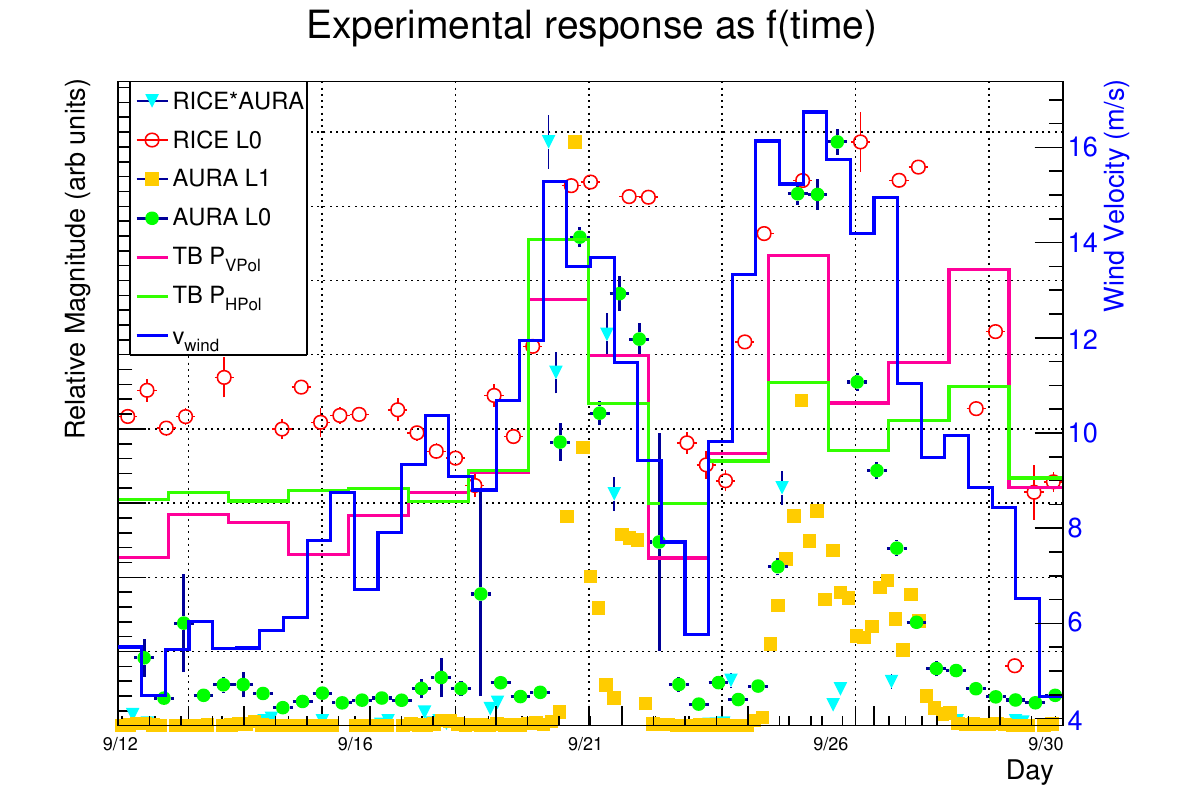}}\caption{Wind velocity distribution (blue line) for September, 2011, superimposed on indicated trigger rate distributions. Shown are the raw ("L0") RICE trigger rate (red circles), the AURA raw trigger rate (green points), the AURA trigger rate for events triggering multiple DRM's ("L1", shown as orange points), and the rate for events which trigger both RICE and AURA to within $10^{-5}$ seconds. Also overlaid are the testbed average waveform power for both vertically (VPol) and also horizontally (HPol) polarized antennas. In all cases, we observe some correlation with wind velocity, beyond an initial (upwards) threshold-crossing of approximately 10 m/s, although the degree of correlation varies event-to-event. 
\label{fig:Sept2011triggerrates}}
\end{figure}

\autoref{fig:Sept2011triggerrates} presents the wind velocity distribution, the raw RICE and AURA trigger rates (``L0''), the RICE x AURA L0 coincidence rate (defined as two triggers with a time stamp within 10 microseconds), and the AURA rate for events having four channels with signals exceeding 6$\sigma_V$ in that channel (``L1''). The ARA station A0 testbed provides additional information, namely, summed signal waveform total power ($\Sigma(V_i^2)$ for the eight vertically polarized (VPol) antennas vs.\ the eight horizontally polarized (HPol) antennas. Several features are evident from this Figure:
\begin{itemize}
    \item The raw RICE trigger rate follows the wind velocity distribution extremely closely, with no evident hysteresis. The enhancement in trigger rate is approximately a factor of two, although this is limited by the limited throughput of the RICE data acquisition system.
    \item Employing a much more efficient, and higher-bandwidth DAQ, the AURA raw trigger rate also tracks the wind velocity distribution, and shows an enhancement of 10--15 over ambient, with a turn-on threshold of approximately 11 m/s
    \item The RICE x AURA coincidence rate also tracks the wind-velocity distribution, indicating that the same (presumably high-amplitude) tribo-electric events observed by one experiment also triggers the other.
    \item The testbed (ARA station 0) similarly shows waveform power characteristics, in both polarizations, which closely track the local wind velocity profile and with a nominal threshold $\approx$10 m/s.
\end{itemize}

\begin{figure}
\centerline{\includegraphics[width=0.9\textwidth]{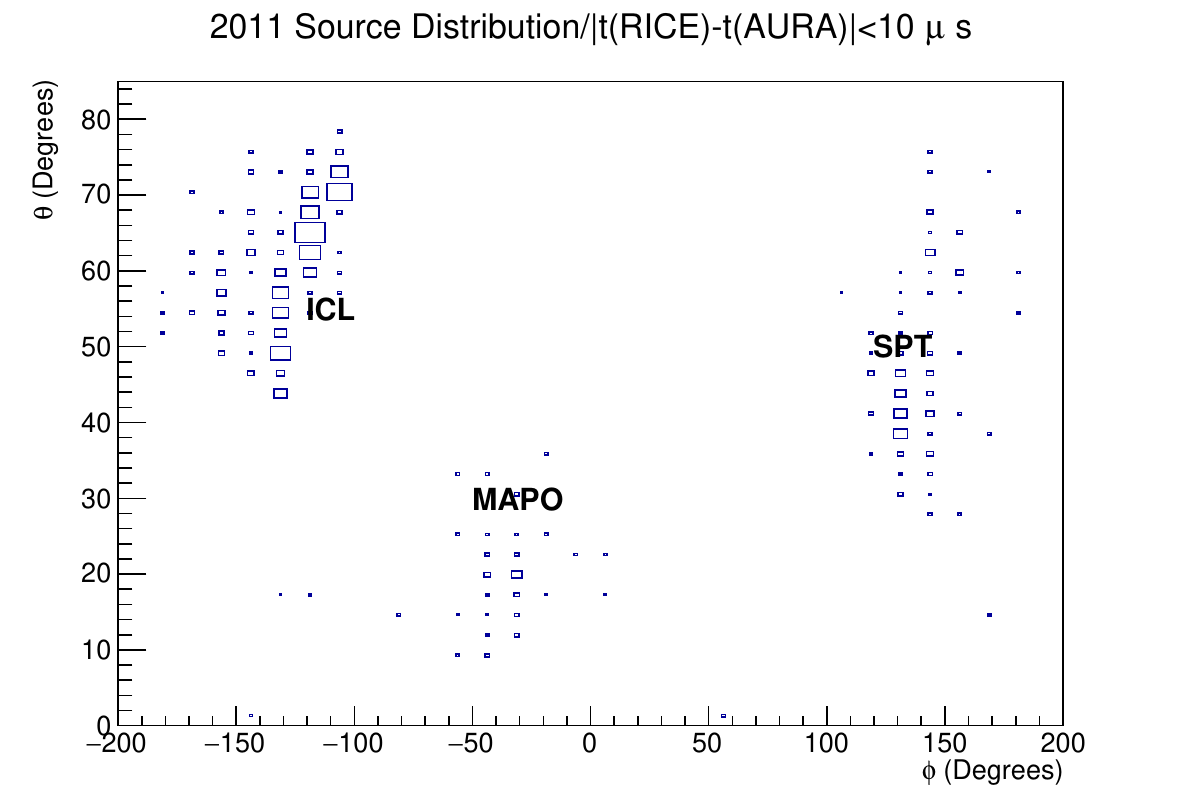}}
\caption{Arrival directions of events recorded during $v_{\mathrm{wind}} > $ 12 m/s - coincidences in the 2011 data samples between RICE, AURA, and ARA. Source reconstruction algorithm finds the three-dimensional space point most consistent with the relative channel-to-channel hit timing, within a given event.}
\label{fig:RxASourcePhiThe}
\end{figure}

Thus, whereas the RICE trigger rate closely follows the wind velocity distribution, the A0 (testbed) rate correlation is only noticeable for time periods when the wind velocity exceeded 15 m/s. These results are consistent with a model where the observed background depends on the distance to the closest above-surface possible discharge site, particularly given the somewhat higher $6\sigma_{\mathrm{SNR}}$ RICE trigger threshold compared to the $\sim 4\sigma_{\mathrm{SNR}}$ ARA trigger threshold.
Although the A0 clock was not well-synchronized to UTC, the sample for which the RICE and AURA trigger times were coincident to within 10 microseconds is approximately background-free. \autoref{fig:RxASourcePhiThe} shows the source distribution for such coincident events; since the AURA receiver antennas are all aligned vertically, receiver signals on the same string do not provide azimuthal source location information.

\subsection{RNO-G}
Similarly to the Antarctic experiments, the RNO-G experiment \cite{Aguilar:2020xnc} is currently under construction in the Northern Hemisphere at the highest point on the Greenland ice sheet (``Summit Camp''). In the summer of 2021, the initial three stations of the planned 35-station array were deployed at 0.9--2.1 km distances from the main station (``Big House''). The RNO-G station design combines both deep antennas (VPol based on the fat dipole RICE design and HPol based on the ARA design) deployed to 100 meter depth and also surface antennas (LPDA's identical to those used by ARIANNA) \cite{Aguilar:2020xnc}. Unlike ARA, which comprises 8 VPol antennas interspersed with 8 HPol antennas over a 20 m x 20 m x 20 m cuboid, the RNO-G geometry emphasizes low trigger threshold via a dedicated phased-array string modeled on the ARA Phased Array design \cite{allison2019design}, plus two `helper' strings to provide azimuthal directional reconstruction. 

The new RNO-G data allow us to investigate correlation of experimental observables with wind velocity at a Northern Hemisphere site, and compare to measurements at South Pole. The two sites are at approximately equivalent elevation, although the humidity and surface temperature at Summit Station typically exceed the corresponding values at South Pole, which may impact the likelihood of triboelectric discharges. We also note that the average surface wind velocities at Summit Station are significantly higher than at South Pole, such that the possibility of wind-related backgrounds, as well as the attractiveness of wind turbines for power provision, are also both correspondingly higher. 

\begin{figure}
    \centering
    \includegraphics[width=1.0\textwidth]{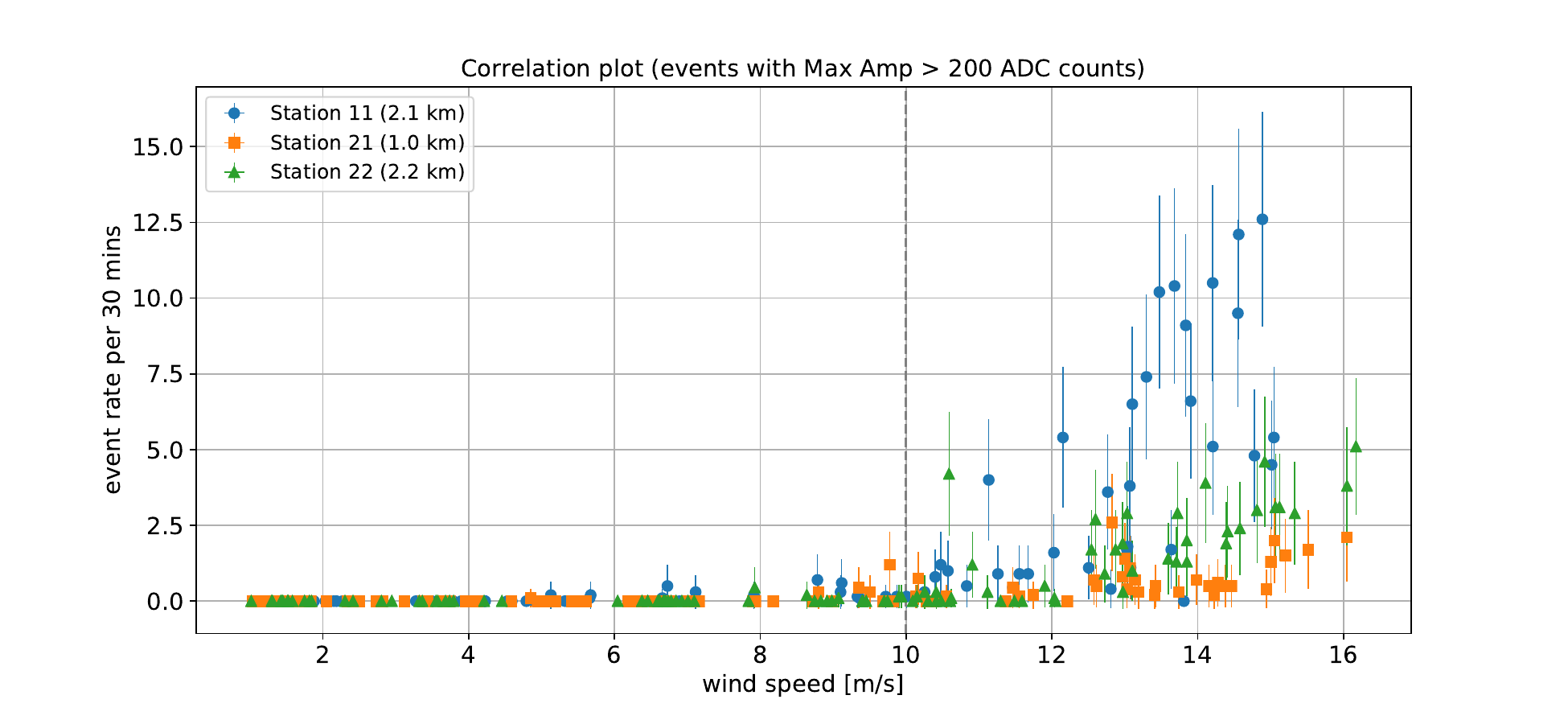}
    \caption{Average RNO-G trigger rate for events containing high-amplitude signals, as a function of average wind speed at the time-of-trigger. Data are shown for the three stations deployed and commissioned in the summer of 2021. The distance to the main building of Summit Station is indicated in the key.}
    \label{fig:RNO-G_wind1}
\end{figure}

Shortly after initial commissioning of the first three stations during the summer of 2021, winds attained velocities of 14 m/s on August 14, 2021 and again on August 28, 2021. Over a three-day period in mid-September, 2021, sustained wind velocities exceeding 14 m/s were again recorded; our study below presents results based on the September, 2021 data sample. As before, we investigate the shape, duration, and frequency and polarization content of candidate triboelectric discharges. As before, we use raw trigger rates as the basic diagnostic of wind correlations. \autoref{fig:RNO-G_wind1} shows the event rate, selecting only those triggered events for which the maximum amplitude signal registered on an individual channel (i.e., antenna) in a given event exceeds 200 ADC counts. This simple requirement selects only extremely large events (SNR = $V_{\mathrm{peak}}/V_{\mathrm{rms}}> 15$). It suppresses contributions from thermal noise triggers and potential self-induced trigger contributions in this preliminary RNO-G data analysis. It results in a low-wind velocity ambient baseline consistent with zero. 

As before, we observe a `turn-on' behavior with a threshold of approximately 10 m/s. However, the increase in event rate is not most evident in the channel closest to the main Summit Station building, suggesting that either a) the discharge site is a more local surface structure, such as the solar panel supports on the surface above the array, b) the experimental trigger threshold of Station 21 is simply higher than Stations 11 and 22 (as of now, the trigger thresholds of the three stations have not been internally calibrated against one another), or c) the correlation with above-surface structures indicated in the RICE data is not universal.

\begin{figure}
\centerline{\includegraphics[width=0.9\textwidth]{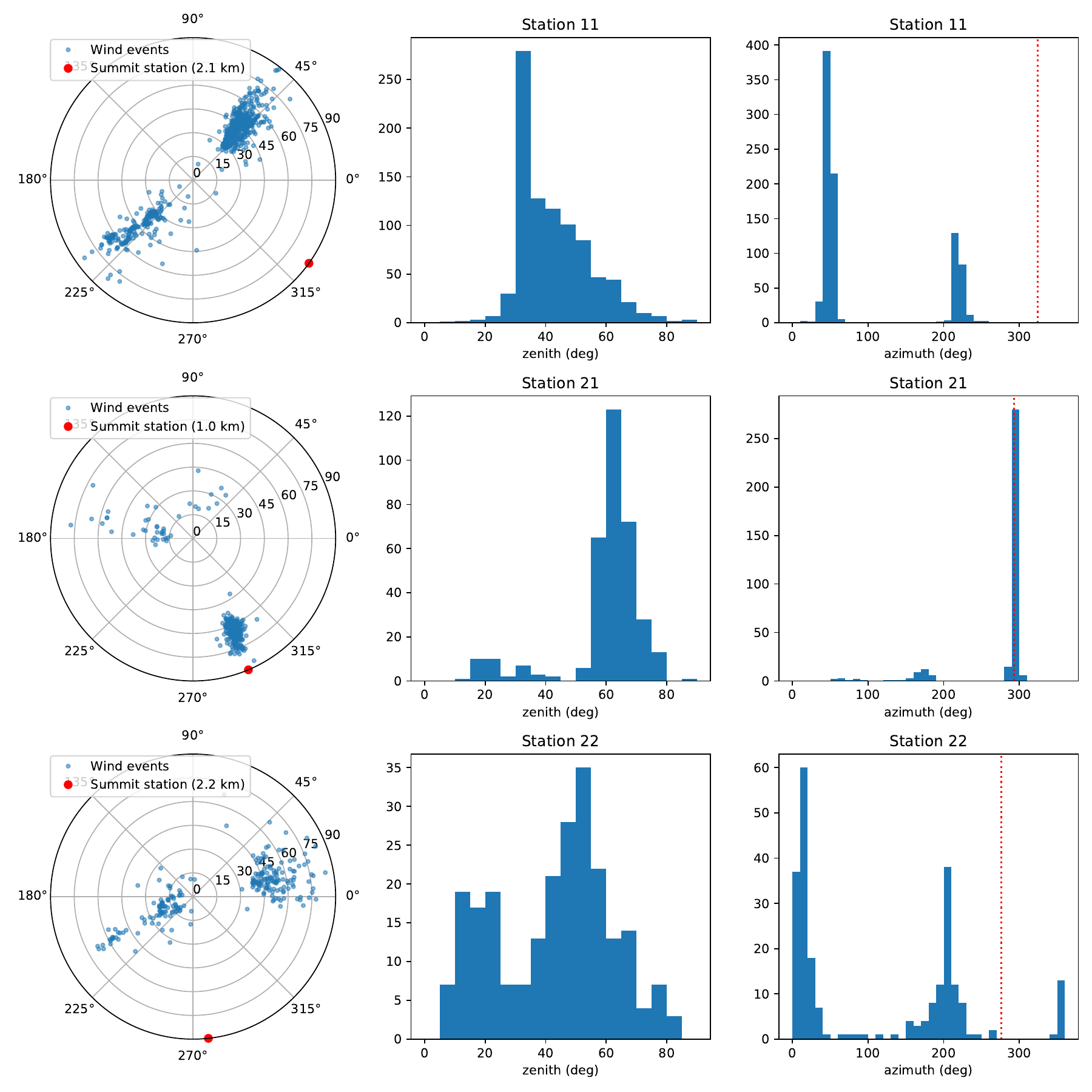}}
\caption{Reconstructed source map for same data set presented in \autoref{fig:RNO-G_wind1} ($v_{\mathrm{wind}} > 10$ m/s, plane-wave) for the three deployed RNO-G stations (11, 21, and 22). Left column: Reconstructed azimuth and zenith angle of incoming direction, with 90$^{\circ}$ in azimuth being North and 0$^\circ$ in radial unit being up. Right: Histograms of the reconstructed angles. The respective direction of main Summit Station is indicated in each map (red marker) and azimuth distribution (dashed line).}
\label{fig:RNO_wind2}
\end{figure}

Although the data have not been fully calibrated, rough source location maps can be constructed from the sample of RNO-G triggers which satisfy the minimum ADC count event selection requirement. \autoref{fig:RNO_wind2} presents a map of both elevation ($\theta$) and azimuth ($\phi$) for this candidate event sample, with the azimuthal direction to the main Summit Station building superimposed. In the case of the closest station and for one particular day, we observe an evident correlation with the main station azimuth. In the other cases, we observe an anisotropic source map with no clear correlation with known structures, although the possibility that there are above-surface structures in the reconstructed source directions, such as the local solar panels, is still being assessed.

Due to the very early nature of the RNO-G data, taken during commissioning of the first stations, a future study with more data will be needed to determine whether there is a distribution of lower amplitude events correlated with increased wind-speeds. Correlations will be quantified with respect to arrival directions, above surface structures and the efficacy of various vetoing strategies. 

\subsection{TAROGE-M}
Although production of tau neutrinos at astrophysical sources is highly unlikely, mixing between neutrino species should result in a roughly equal ratio of $\nu_e:\nu_\mu:\nu_\tau$ astrophysical neutrinos, as measured at Earth. The high mass, and therefore short lifetime of the tau lepton therefore affords a unique detection strategy to elevated ($h>$1 km) radio arrays - namely, searches for radio-frequency emissions resulting from conversion of tau neutrinos below in terrestrial rock. This strategy has been adopted by the TAROGE-M experiment, sited atop Mt. Melbourne, Antarctica at an elevation of 2720 m.

Consisting of an elevated array of custom log-period dipole antennas similar to those used by ARIANNA, the TAROGE-M experiment has recently conducted a search for radio emissions from ultra-high energy cosmic rays, based on 26.55 days of livetime accumulated in February, 2020.  Use of a drone pulser transmitter allows calibration of the trigger efficiency over 2$\pi$ solid angle, with a measured threshold $SNR\approx 4$. TAROGE-M finds that approximately 99.9\% of their 1257122 total triggers are collected when the winds are highest (with velocities exceeding 7 m/s and are readily separated from seven high-quality UHECR candidates, using a simple cut on wind velocity and temporal clustering. Similar to other experiments, TAROGE-M finds that the power spectra of their wind events are shifted to the lower edge of their band. Moreover, they find that the highest amplitude signal is recorded in the antenna closest to the primary experimental surface structure, consistent with a nearby metallic discharge model. The noticeable gradient in recorded signal voltages, across the TAROGE-M array, also support a nearby, local source. 

\begin{figure}
    \centering
        \includegraphics[width=\textwidth]{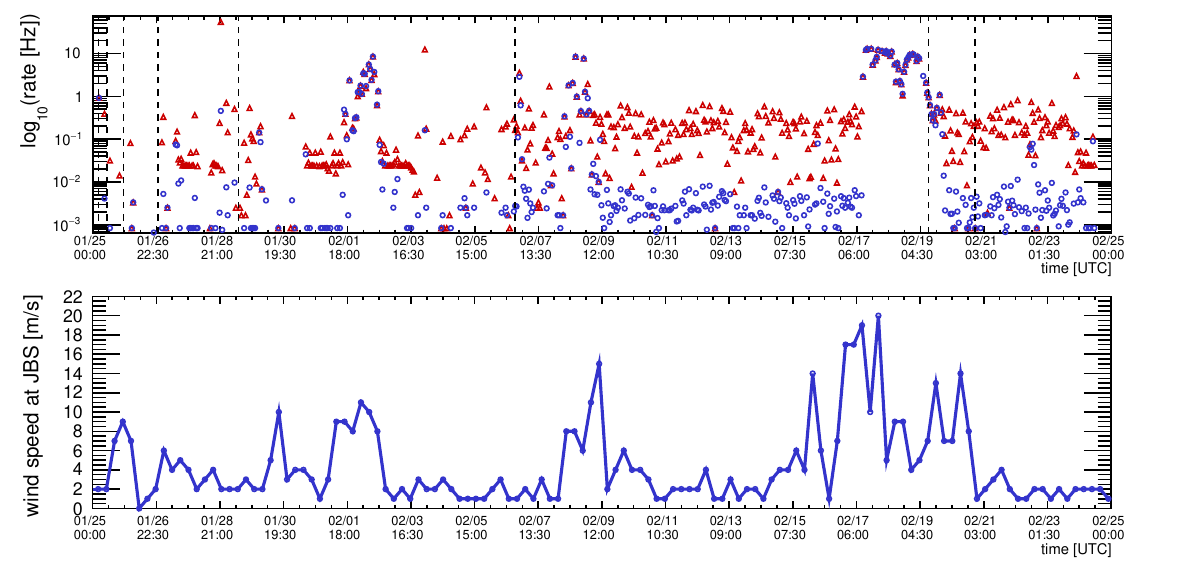}
    \caption{ 
       Top: TAROGE-M event rate (Hz, with suppressed zero) over time for channel-level (red) and station-level (blue) L1 trigger.
       %
       The vertical dashed lines mark the starting time of new run periods with slightly different trigger thresholds.
       Bottom: wind speed data recorded at the Jang Bogo station (JBS) concurrently with TAROGE-M operation.
    }
      \label{fig:wind_rate}
    \end{figure}

The correlation of wind speed with trigger rate, as a function of time, is shown in Figure \ref{fig:wind_rate}, illustrating the dependence of the station-level trigger on local wind velocity.
Interestingly, damage at the input to their front-end low-noise amplifiers was noted following one of the high-wind periods, suggesting discharge within the DAQ itself.


\section{Summary and Conclusions}
We summarize our findings as follows:
\begin{enumerate}
\item All radio neutrino experiments have observed radio-frequency signals correlated with high wind velocities exceeding 10 m/s, at South Pole, Antarctica and Summit Station, Greenland. The precise threshold may depend on the exact antenna and trigger configuration.
These observations are consistent with data taken at other locales within the last century indicating electric fields associated with wind blowing over water/snow/ice, with an experiment-specific discharge threshold of order 10 m/s. This suggests an initial experimental mitigation strategy would likely reject neutrino background events which trigger during high-wind times and/or temporally cluster.  
At the site anticipated for the IceCube-Gen2 radio experiment, for example, a neutrino search strategy which eliminates all data taken when wind velocities exceed 10 m/s preserves $\approx 95$\% of the candidate neutrino search data sample. Although the coherence distance for wind velocities (both magnitude and direction) has not been measured at South Pole, this study indicates that incorporating anemometers into the baseline IceCube-Gen2 experimental design may provide important information for future neutrino and cosmic ray searches.
\item Although the observed signals have durations that can extend for a microsecond, they may have sufficiently sharp leading edges to permit source reconstruction, and therefore identification as background. Such events may, in some cases, be difficult to distinguish from the expected impulsive signal waveforms resulting from both ultra-high energy neutrinos in-ice, as well as ultra-high energy cosmic rays in-air.  
\item In addition to potentially saturating the throughput of the data acquisition systems, high-wind events are typically characterized by extended signal waveforms (rather than the narrow, impulsive signals expected from a true in-ice neutrino interaction) with power preferentially shifted to lower frequencies (f$<$250 MHz; see \autoref{fig:windvPwr}, for example).
\item Observed signals, presumably resulting from triboelectric coronal discharge, have significant power in both vertical as well as horizontal polarization, consistent with random discharge geometries. 
\item High-wind events are typically evidenced by an overall enhancement in trigger rates for the polar experiments discussed herein, potentially adding ${\cal O}(10^{4-6})$ excess triggers per year, depending on the experiment and local wind conditions. Experience from ARA indicates that, out of that excess ${\cal O}$(1 M) excess triggers, a handful of events per year, observed during high wind-periods, may evade standard background suppression strategies and also have durations shorter than the typical temporal response of the detector. 
\item For some experiments, reconstruction of surface source locations indicate a clear preference for electric field discharge on nearby metallic structures (the `lightning rod' effect), favoring a model with significant airborne charge and disfavoring a model where charge build-up is concentrated solely on the snow surface. However, the possibility of lower-amplitude, more isotropic source locations cannot be ruled out by the present data.
\item For buried antenna arrays ($>{\cal O}(10~m)$ in depth), surface-originating noise associated with the triboelectric effect may be rejected given the sub-degree polar angle source resolution of such arrays. For typical neutrino searches, such a geometric cut corresponds to a worst-case loss of approximately 20--25\% in sensitive neutrino volume.
\item Rejection of wind-generated backgrounds to radio signals arising from ultra-high energy cosmic ray interactions in-air are less readily suppressed using source reconstruction strategies due to their common near-surface origin. However, previous ARIANNA analysis demonstrates the poor correlation of event waveforms with signal shapes expected for UHECR and novel rejection strategies based on machine learning approaches have already demonstrated considerable power and promise.
\item Data drawn from multiple experiments indicates that peak radio emissions are observed for wind velocities of approximately 15 m/s; at higher wind velocities, effects appear to mitigate, possibly owing to the difficulties of surface charges to conglomerate.  
\end{enumerate}
Overall, triboelectric effects will almost certainly continue to trigger radio-sensitive experiments seeking detection of ultra-high energy cosmic rays. As enumerated above, mitigation strategies will likely combine local wind velocity data, source reconstruction characteristics (originating on the surface or near identifiable surface structures), frequency content (power spectra shifted to lower frequencies), and goodness-of-fit to neutrino templates. To maximize sensitivity to both radio emissions from ultra-high energy cosmic rays (UHECR) as well as ultra-high energy neutrinos (UHEN), efforts to develop additional source identification algorithms and strategies seem warranted, although given the observed variability, background rejection schemes will likely have to be tailored individually for a given experiment.

\section*{Acknowledgments} 
For this article information and data from all previous radio neutrino experiments were compiled. We thank all involved experimental groups for agreeing to collaborate on this broader topic highly relevant for the experimental future. We in particular acknowledge dedicated data analysis for this paper by M.~Mikhailova (AURA, RICE), D.~Besson (SATRA, RICE, ARA), L.~Zhao (ARIANNA), and S.~Bouma and M.~Cataldo (RNO-G). 

We gratefully acknowledge the support of all funding agencies that supported these experiments, in particular the National Science Foundation, over the term during which these data were taken. Also, none of the experiments would have been possible without the invaluable field support of various agencies and staff at South Pole, McMurdo, and Summit Station.

The ARIANNA collaboration acknowledges the U.S. National Science Foundation-Office of Polar Programs, the U.S. National Science Foundation-Physics Division (grant NSF-1607719) and NSF grant NRT 1633631. C.~Glaser acknowledges funding from the German research foundation (DFG, grants GL 914/1-1). Support is acknowledged from the Taiwan Ministry of Science and Technology, and the Swedish Government strategic program Stand Up for Energy. 

The ARA collaboration acknowledges the National Science Foundation (NSF) Office of Polar Programs and Physics Division for funding support, as well as the Taiwan National Science Councils Vanguard Program NSC 92-2628-M-002-09 and the Belgian F.R.S.-FNRS Grant 4.4508.01. K.~Hughes thanks the NSF for support through the Graduate Research Fellowship Program Award DGE-1746045. B.~A.~Clark thanks the NSF for support through the Astronomy and Astrophysics Postdoctoral Fellowship under Award 1903885. A. Connolly thanks the NSF for Award 1806923. S.~A.~Wissel thanks the NSF for support through CAREER Award 2033500. A.~Vieregg thanks the Sloan Foundation and the Research Corporation for Science Advancement, the Research Computing Center and the Kavli Institute for Cosmological Physics at the University of Chicago for the resources they provided.  R.~Nichol thanks the Leverhulme Trust for their support. D.~Z.~Besson, I.~Kravchenko, D.~Seckel and D.~Williams thank the National Science Foundation for their generous support of the IceCube EPSCoR Initiative (Award ID 2019597).

The RNO-G collaboration acknowledges the Belgian Funds for Scientific Research (FRS-FNRS and FWO) and the FWO programme for International Research Infrastructure (IRI), the National Science Foundation through the NSF Awards IDs 2118315, 2112352, 211232, 2111410 and the IceCube EPSCoR Initiative (Award ID 2019597), the German research foundation (DFG, Grant NE 2031/2-1), the Helmholtz Association (Initiative and Networking Fund, W2/W3 Program), the University of Chicago Research Computing Center, and the European Research Council under the European Unions Horizon 2020 research and innovation programme (grant agreement No 805486).

\bibliography{a.bib}
\bibliographystyle{JHEP} %

\end{document}